\documentclass{aa}  

\usepackage{graphicx}
\usepackage{booktabs}
\usepackage{multirow}
\usepackage[hidelinks]{hyperref}
\usepackage{academicons}
\usepackage{xcolor}
\definecolor{orcidlogocol}{HTML}{A6CE39}
\usepackage[switch]{lineno}

\usepackage{mathcomp}
\usepackage{txfonts}

\usepackage{subcaption}         
\usepackage{lscape}             
\usepackage{placeins}           

\begin{document} 
   \title{Observing Double White Dwarfs with the Lunar GW Antenna}

   \author{G. Benetti \inst{1}
   \and
   M. Branchesi \inst{2,3}
   \and
   J. Harms \inst{2,3}
   \and
   J. P. Zendri \inst{4}
}
   \institute{Dipartimento di Fisica e Astronomia Galileo Galilei, Università di Padova,
 35131 Padova, Italy\\ \email{giovanni.benetti.1@studenti.unipd.it}
              \and 
              Gran Sasso Science Institute (GSSI), I-67100 L'Aquila, Italy
              \and 
              INFN, Laboratori Nazionali del Gran Sasso, I-67100 Assergi, Italy
              \and
              INFN, Sezione di Padova, Via Marzolo 8 Padova, Italy
             }

 
  \abstract
   {The Lunar Gravitational Wave Antenna (LGWA) is a proposed gravitational-wave detector that will observe in the decihertz (dHz) frequency region. In this band, binary white dwarf systems are expected to merge, emitting gravitational waves. Detecting this emission opens new perspectives for understanding the Type Ia supernova progenitors and for investigating dense matter physics.}
   {In this paper, we present the capabilities of LGWA to detect and localize short-period double white dwarfs in terms of sky locations and distances. The analysis employs realistic spatial distributions and merger rates, as well as binary-mass distributions informed by population-synthesis models.}
   {The simulated population of double white dwarfs is generated using the \textsc{SeBa} stellar-evolution code, coupled with dedicated sampling algorithms. The performance of the LGWA detector, both in terms of signal detectability and parameter estimation, is assessed using standard gravitational-wave data analysis techniques, including Fisher matrix methods, as implemented in the \textsc{GWFish} and \textsc{Legwork} codes.}
   {The analysis indicates that, over 10 years of observation, LGWA could detect approximately
30 monochromatic Galactic sources and 10 extragalactic mergers, demonstrating the unique potential of decihertz gravitational-wave detectors to access and characterize extragalactic DWD populations. This will open new avenues for understanding Type Ia supernova progenitors and the physics of DWDs.}
   {}


   \maketitle
   \nolinenumbers
%

\section{Introduction}
\label{sec:intro}

The Lunar Gravitational-Wave Antenna (LGWA) is a proposed Moon-based gravitational-wave (GW) detector with an observation band spanning from about 1\,mHz to 1\,Hz \citep{Ajith:2024mie, Harms_2021}. The working principle relies on the excitation of the Moon's vibrational modes, in particular the quadrupolar eigenmodes, induced by the apparent tidal forces associated with the transit of a GW. The amplitude of the modes can be measured by an array of four seismometers \citep{vanHeijningen:2023esw}, from which it is possible to reject the seismic background and reconstruct the original waveform \citep{Harms_2022}. The possible observation band is determined by both the lunar response to GW excitation and the instrumental detection noise \citep{Cozzumbo:2023gzs}. The LGWA can fill the gap in the dHz band, bridging between the space-detector LISA and future terrestrial GW detectors like the proposed Einstein Telescope \citep{Punturo:2010zz, Branchesi:2023mws} and Cosmic Explorer \citep{Evans:2021gyd, Evans:2023euw, Gupta:2023lga} in a very physics-rich frequency window \citep{Ajith:2024mie, Sedda:2021yhn}.

Other proposed lunar-based GW detectors are the Laser Interferometer Lunar Antenna (LILA), and the Laser Interferometer On the mooN (LION), which all consist of triangle-shaped laser interferometers \citep{Jani:2020gnz, LILA, Amaro-Seoane:2020ahu}. In the dHz band, other proposed detectors are the Deci-hertz Interferometer Gravitational wave Observatory (DECIGO) with its precursor mission B-DECIGO \citep{Kawamura:2006up, Kawamura:2011zz, Sato:2017dkf, Isoyama:2018rjb}, the Taiji, Tianqin and TianGO detectors \citep{Luo:2021qji, Hu:2017mde, TianQin:2015yph, Kuns:2019upi}, the Big Bang Observer (BBO) and the Advanced Laser Interferometer Antenna (ALIA) as follow on missions to LISA \citep{Crowder:2005nr}. All the aforementioned, proposed dHz detectors are long-baseline laser interferometers.

White Dwarfs (WD) are the evolutionary endpoint of stars with initial mass $M\lesssim 10 \, M_\odot$  \citep{Saumon:2022gtu}; they are compact objects (with high density, $ \sim 10^6 \, \text{g/cm}^3$) supported by the electron degeneracy pressure, which can sustain a stable configuration only for $M<M_\text{Ch}$, where $M_\text{Ch}\approx 1.4 M_\odot$ is the Chandrasekhar mass \citep{Chandrasekhar:1931ih}. WDs can be found in binary systems, due to the individual evolution of the component stars. The processes that lead to the formation of a short-period DWD that will merge in less than a Hubble time are complex and involve at least two episodes of mass transfer, one of which is a common envelope (CE) phase \citep{Woods:2011dd}, during which both stars are embedded in a gas envelope resulting from the expansion of one of the two stars in the latter stage of evolution. The envelope exerts a drag force on the system, which consequently shrinks; the energy lost by drag contributes to heating the CE, which is eventually expelled, leaving a short period DWD. The binary shrinks due to the emission of GWs and ultimately merges. A merger between two WDs with a total mass $M_\text{tot}>M_\text{Ch}$ is a potential progenitor of Type Ia supernovae (SN Ia) \citep{Webbink:1984ti, WANG_prog_Ia}.
A SN Ia is a thermonuclear explosion believed to be generated by the accretion of a Carbon-Oxygen (CO) WD from a companion; when the Chandrasekhar mass is reached, the star collapses due to the insufficient electron degeneracy pressure, starting a carbon-burning phase that produces heavier elements up to iron. This sudden energy production results in an explosion that incinerates the star, probably leaving no further remnants. 

The scenario involving two CO WD is called double-degenerate (DD), contrary to the single-degenerate scenario (SD) in which the supernova results from the He and H accretion of a CO WD from a non-degenerate companion. We note that both channels are probably present, especially taking into account the phenomenological variety of SN Ia events. Both scenarios benefit from indirect evidence, even if the DD scenario would better explain the lack of H and He emission in SN Ia spectra \citep{SNIa_Hspectra}. In conclusion, there is still no definitive evidence of the progenitor's nature \citep{Ruiter:2024kak, Maoz:2011iv}. The quest for the progenitor is particularly important given the role of SN Ia as standard candles in the measurement of the Hubble constant \citep{Riess_H0_SNIa, Riess:2021jrx, Pascale:2024qjr}, and the tension with CMB measurements \citep{Planck:2015fie, Planck:2018vyg, DiValentino:2021izs}. 

LGWA offers a novel way to characterize short-period DWDs in terms of properties and rate. By comparing these measurements with the observed SNIa rate, LGWA will enable the identification of the dominant formation channel and progenitor systems of SNe Ia. During the merging of DWD systems the GW carries information regarding the matter effects that shape the interaction dynamics. LGWA observations will probe the physical processes involved in the last years of evolution of the DWDs, exploring processes that are only observable with detectors sensitive in the dHz band.

Observing GWs from extragalactic DWDs would provide a means of independently calibrating cosmic distances. By identifying the host galaxy and combining its redshift with the luminosity distance inferred from the GW signal, it becomes possible to estimate the Hubble constant. These systems can act as standard sirens \citep{Maselli:2019mzt, Schutz:1986gp, Krolak:1987ofj}, offering a novel and complementary approach to addressing the Hubble tension, similar to the method employed with GW170817 \citep{Schutz:1986gp, LIGOScientific:2017adf}. 
A multimessenger detection of a DWD that bursts into a SN Ia would directly prove the presence of the DD formation channel and calibrate the SN Ia as standard candles. The opposite (GW detection without EM counterpart or EM detection without GW event) does not necessarily rule out the possibility of a DD event, as there are models in which the deflagration happens a long time after the coalescence of the two CO WD \citep{SNIa_delayed}. Furthermore, the detection of a GW signal without an electromagnetic counterpart does not preclude the identification of the host galaxy. If the DWD is well-localized by the GW signal, it can enable a "dark siren" measurement of the Hubble constant \citep{Finke:2021aom, Leandro:2021qlc, Gair:2022zsa}.

While lower-frequency instruments like LISA are expected to detect approximately 10,000 to 20,000 \citep{Korol:2017qcx} or more \citep{Korol:2021pun} resolvable inspiral signals from DWD systems in the Milky Way, and millions more contributing to a confusion-limited GW foreground at low frequencies \citep{Korol:2021pun}, LGWA will be uniquely capable of observing the final inspiral phase and merger of extragalactic DWDs. In fact, the merging frequency for these systems matches the peak sensitivity of LGWA (sec.~\ref{sec:1_DWD_general}). A simplified estimate of the detector's observational capability can be obtained by comparing the GW amplitude from a nearby DWD source with the expected power spectral density (PSD) of the instrument, as detailed in Appendix  \ref{sec:plain_analysis}. A similar method was adopted by \cite{Marcano:2025uct} in their analysis of LILA (former GLOC) and LGWA. However, assessing the realistic potential of LGWA requires more than analyzing a single system with fixed parameter. To reliably estimate detection rates and parameter estimation capabilities, these results must be placed within the context of a realistic, population-wide model of DWDs. In particular, the binary-mass distribution is highly non-trivial (see Sect.~\ref{sec:population}), shaped by complex stellar evolution processes and diverse star-formation histories. Assuming a simplified, approximately uniform mass distribution leads to a significant overestimation of the number of detectable systems.

In this work, an efficient and reliable method is provided to simulate the rare but loud population of short-period DWDs (Sect.~\ref{sec:population}), taking into account the absolute abundance weighted by the supernovae type Ia (SNIa) rate, and the detailed distribution of the DWD set in the parameter space arising from the main astrophysical phenomena. The simulated population is then analyzed in order to assess the LGWA performances in terms of detection and parameter estimation, with particular care with regards to the localization capabilities (sec.~\ref{sec:results}). Our analysis, based on the assumption that observed SN are produced by DWD merger, suggest that LGWA will be able to detect from a few to ten extragalactic events during the ten-year mission lifetime, however the uncertainty on this estimation is heavily affected by the merging frequency of the systems, which depends on the complex physics of the matter-dominated merging process.

Throughout this paper, $H_0 = 70 \, \text{km s}^{-1}\text{Mpc}^{-1}$ is assumed.

\section{DWD merging process}
\label{sec:1_DWD_general}

\subsection{Merging frequency for DWD systems}
\label{subsec:merging_frequency}

The small radius of WDs (order of magnitude of $10^6$\,m) implies that the GW-radiating phase is very long, before other effects such as magnetic braking, Roche overflow and tidal disruption take over. In the following, the maximum GW frequency reached by a DWD system as a function of the WD masses is estimated considering different masses $m_1$, $m_2$, eccentricity $\varepsilon=0$ and WD radius $R_\text{WD}$. The assumption $\varepsilon=0$ is justified by the fact that close WD binaries exhibit highly circularized orbits, a consequence of the CE phase experienced during the evolution. Hydrodynamic simulations \citep{DWD_hydro} show that the merging occurs when the distance between the WDs is approximately two to three times the WD radius. This limit is heavily dependent on the physics of the merging process, and thus is not easily to model with simple considerations. In the following, we consider two scenarios: the ``Roche scenario'' and the ``Contact scenario''. The estimates corresponding to the Roche and Contact approximations represent the two extreme scenarios that encompass the range of realistic detection situations, which will become accessible as waveform models improve.

\paragraph{Roche scenario} 
The matter of two WD behaves as an inviscid fluid; the tidal disruption happens when the least massive star fills its Roche lobe. This limit is set by the condition that the Roche lobe volume shall be equal to the star volume. The Roche volume is calculated  following the Roche radius prescription by \cite{Eggleton:1983rx}, where $q = m_1/m_2$ is the mass ratio:
\begin{equation}
    r_R\approx d\frac{0.49q^{2/3}}{0.6q^{2/3}+\ln{(1+q^{1/3})}}
\end{equation}
The WD volume is given by the theoretical mass-radius (MR) relation \citep{Padmanabhan}:
\begin{equation}
    R(M) \approx \frac{0.022}{\mu_e}\bigg(\frac{M}{M_{Ch}}\bigg)^{-1/3}\Bigg[1-\bigg(\frac{M}{M_{Ch}}\bigg)^{4/3} \Bigg]^{1/2}R_\odot
    \label{eq:mass-radius}
\end{equation}
The MR relation can also be simulated by the population synthesis code \textsc{SeBa} \citep{SeBa_1,SeBa_2} to better account for low-density regime effects \footnote{An accurate description of the \textsc{SeBa} usage is provided in Sect.~\ref{subsec:SeBa_and_convolution}, for now it is used only to simulate the MR relation.}. Compared to the more complete \textsc{SeBa} MR relation, we verify that the simple theoretical prediction (eq. \ref{eq:mass-radius}) with $\mu_e=2$ holds well (better than 1\%) for masses $\gtrsim 0.5 M_\odot$; at lower masses the complete degeneration approximation is not satisfied, making eq. \ref{eq:mass-radius} less reliable. However low mass systems are not relevant, due to the excessively low merging frequency. In the following eq. \ref{eq:mass-radius} will thus be used to approximate the MR relation. Note that the beginning of the Roche overflow is set by the least massive WD, and the GW waveform is not sharply cut at this frequency. 

\paragraph{Contact scenario}
The matter of two WD behaves as a rigid solid; the disruption happens at the contact of the two spherical WDs. In this case, the merging distance is given by $d = R(m_1)+ R(m_2)$, with $R(M)$ from eq. \ref{eq:mass-radius}. This assumption brings the risk of overestimating the signal-to-noise-ratio (S/N), however it represents a much more realistic assumption: the frequency cutoff for the Roche scenario is to be intended as a strongly conservative estimation of the maximum frequency, up to which the waveform approximants available today are reliable. In reality we can expect the waveform to reach and pass the contact frequency, even if strongly influenced by matter effects. To account for the range between these two extreme scenarios, in this paper we consider both the scenarios for the extragalactic DWD population. For the Galactic population only the Roche scenario is applied, since it is already largely sufficient to assure a detection.

\paragraph{} In the Newtonian limit, the orbital frequency is
\begin{equation}
\label{eq:freq_binary}
    \Omega^2=(2\pi f_\text{orb})^2=\frac{GM}{d^3}
\end{equation}
The frequency of the GW quadrupole mode is $f=2f_\text{orb}$.

\begin{figure}
    \centering
   
    \includegraphics[width=\linewidth]{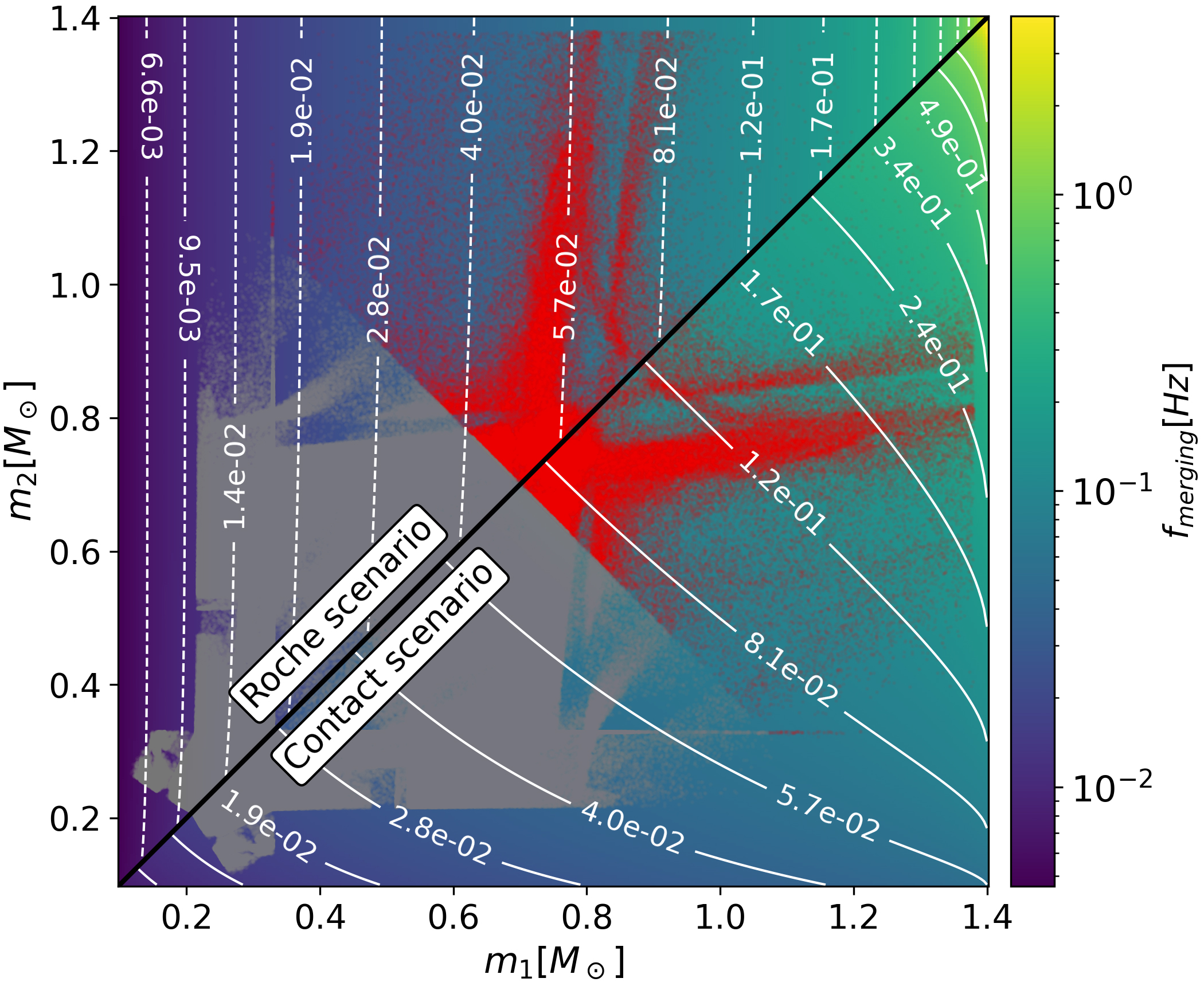}
    \caption{Dependence of the merging frequency with respect to the component masses for the Roche scenario (upper semiplane) and contact scenario (lower semiplane). The reported population (red for $m_1+m_2>M_\text{Ch}$ and gray for $m_1+m_2<M_\text{Ch}$) is presented in Sect.~\ref{sec:population}, and corresponds to the simulated Milky Way DWD population.}
    \label{fig:F1}
\end{figure}

Figure \ref{fig:F1} shows the approximate maximum frequency reached before the merging as a color scale with white contours for some values of $f_\text{max}$. The Roche and contact scenarios are represented by the upper and the lower half-planes, respectively. The simulated Galactic population that will be discussed in Sect.~\ref{sec:population} is overlaid in the plot, distinguishing the super-Chandrasekhar fraction (red) from the sub-Chandrasekhar fraction (gray). Note that, even if in principle a super-Chandrasekhar DWD can exist with a mass combination of $m_1>0.7\,M_\odot$, $m_2<0.7\, M_\odot$, the population extends mainly in the region where both masses are $>0.7\, M_\odot$. In particular, the most populated region in the high-mass ($M_\text{tot}>M_\text{Ch}$) area is the ''stripe'' defined by $m_1 \in [0.7, 1.4]\,M_\odot$, $m_2 \approx 0.8\,M_\odot$; for simplicity, in the following, this strip is called ''super-Ch branch''. The CO DWDs, namely those with component masses approximately $0.5<M_{CO}/M_\odot< 1.2$, which are the principal candidates for a SN Ia progenitor, merge at $0.05$ - $0.1$\,Hz, namely the most sensitive frequency band of LGWA. At these frequencies, neither LISA nor ET have sufficient sensitivity to detect the DWD merger, making LGWA a uniquely capable instrument for its observation \citep{Ajith:2024mie}. Note that the systems above the super-Ch branch are extremely rare. It is customary to use a binary defined by $m_1 = m_2 = 1.2 M_\odot$ or even $1.4 M_\odot$ to quickly test the capabilities of a detector when considering DWD mergers. However, these results show that such simplified estimates have limited value, as they do not accurately reflect the true underlying mass distribution.

\subsection{The binary system GW emission}
\label{sec:binary_system}

The long observing duration of the LGWA mission, combined with its low-frequency sensitivity, makes it particularly well-suited for detecting long-lived monochromatic sources such as DWDs, for which, given a sufficiently wide orbital separation, the orbital evolution is driven predominantly by GW emission.

The system is modeled by two masses $m_1$ and $m_2$ separated by a distance $d$ following quasi-circular orbits on the $xy$ plane: for DWDs, a treatment of the eccentricity $\varepsilon$ is not necessary (see Sect.~\ref{sec:population}). In the Keplerian limit, the slow inspiral of the two point-masses follows a series of stationary circular orbits with frequency given by eq. \ref{eq:freq_binary}.
The energy radiated from the system via GW emission is \citep{Peters:1963ux}:
\begin{equation}
    \frac{dE}{dt} = - \frac{32}{5}\frac{G^4}{c^5}\frac{m_1^2m_2^2(m_1+m_2)}{d^5}f(\varepsilon)
   \label{eq:binary_energyloss}
\end{equation}

Where $f(\varepsilon)$ an eccentricity correction that reads $f(0)=1$, so it will not be considered in the following.

The energy of the Keplerian system is $E(d)=-\displaystyle\frac{1}{2}\frac{Gm_1m_2}{d}$. So, by substituting $d(E)$ in eq.~\ref{eq:binary_energyloss} and integrating, it is possible to obtain $E(t)$ and thus $d(t)$. Using this result in eq. \ref{eq:freq_binary} results in:
\begin{equation}
     f_{GW}^{-\frac{8}{3}}(t)=f_0^{-\frac{8}{3}}-\frac{256}{5}\frac{G^\frac{5}{3}\mathcal{M}^\frac{5}{3}\pi^\frac{8}{3}}{c^5}\cdot t \equiv f_0^{-\frac{8}{3}}-\Delta(\mathcal{M})\cdot t
     \label{eq:GW_frequency}
\end{equation}
where $f_0$ is the GW frequency at $t=0$, and $\mathcal{M}$ is the chirp mass, $\mathcal{M}=(m_1m_2)^{3/5}(m_1+m_2)^{-1/5}$. The merger time $t_m$ can be obtained by posing $f(t_m)=\infty$, thus\footnote{Note that this expression is obtained in the Newtonian limit, so it is an approximation that is valid only for $d \gg R_\text{Sc}$ ($R_\text{Sc}=2GM/c^2$ being the Schwarzschild radius); this regime is not reached since the merging occurs well before.}
\begin{equation}
\label{eq:merging_time}
    t_m=\frac{f_0^{-\frac{8}{3}}}{\Delta(\mathcal{M})}
\end{equation}

A waveform template is used for the estimation of the detector response, chosen from a set of approximants from \textsc{LALsimulation} (part of \textsc{LALsuite}, \citealt{lalsuite}), a software package currently used by the LIGO/Virgo/KAGRA collaboration. We use the template \textsc{IMRPhenomD} \citep{ Husa:2015iqa, Khan:2015jqa, imrphenomd}. The model is very accurate in the LGWA frequency band and is already implemented in the Fisher-matrix code used for this study, \textsc{GWFish}\footnote{https://gwfish.readthedocs.io/}(see Sect.~\ref{subsec:GWFish}). Furthermore, it enables sufficiently fast analyses, even for large-population studies. In the LGWA frequency band, the modeling is also very accurate. The template depends on a set of parameters: the two masses $m_1$, $m_2$, the luminosity distance, the inclination angle $\theta$, namely the angle between the observer and the total angular momentum of the system, the polarization angle $\psi$, the initial phase of the waveform, the position in the sky in the usual (ra, dec) coordinates, the time of arrival at the geocenter at the nominal merger (without considering the real merger, namely the time at which the Roche overflow occurs), and a frequency cutoff used to simulate both a system that doesn't merge during the observation period, and for which the waveform is truncated, or the occurrence of the Roche overflow in merging systems. A description of these parameters can be found in the documentation of \textsc{GWFish}.

\section{Generation of the synthetic population}
\label{sec:population}
In order to simulate with accuracy the response of LGWA to the DWD population, inside and outside our Galaxy, it is prominent to generate a reliable synthetic population. The adopted procedure is summarized in this section; further details and explanations are reported in Appendix \ref{appendix:methods}.

\subsection{Population size estimation}
\label{subsec:pop_size_estim}
In order to generate a realistic population, it is important to estimate the number of DWD systems in our Galaxy and in the local Universe. Our analysis assumes that every super-Chandrasekhar DWD system leads to a SN Ia, and therefore the SN Ia rate derived from this assumption should be considered a lower limit, since other final outcomes are possible from a super-Chandrasekhar DWD merging \citep{Luo:2024qwa}. However, two opposite effects must be considered too: 
firstly, some SNe Ia may originate from sub-Chandrasekhar mass binaries \citep{SDSS_WD_survey, Ruiter:2024kak, Scalzo:2014wxa, Liu:2023qmw}, and secondly, some of the SNe Ia may result from the SD scenario, in a mixed progenitor scenario, as well as other proposed scenarios \citep{Ruiter:2024kak, Liu:2023qmw}. At this stage, it is not realistically possible to determine how these different contributions balance out. GW observations will also contribute to clarifying this aspect.

The adopted SN Ia rate in our Galaxy is $r=(5.4\pm1.2)\cdot10^{-3}$\,yr$^{-1}$ \citep{SNIa_rate_Li}, and is considered constant over time: this approximation is valid on the timescales of the detectable DWDs (those characterized by a GW frequency above 1\,mHz) because the time required to merge starting from $f_i=1$\,mHz is around 1 to 3\,Myr depending on the masses, sufficiently small compared with the entire Hubble time (13.5 Gyr) to consider the SN Ia rate locally fixed. In the following, we derive the density distribution in frequency space of the DWD population necessary to sustain this rate.

Let $\lambda(f)$ be the DWD spectral density, such that $dN=\lambda(f)df$ is the number of DWDs with frequency between $f$ and $f+df$. The merging of these $dN$ systems will occur in a time interval $dt=t_{m}(f)-t_{m}(f+df)$, where $t_m$ is the merger time given by eq. \ref{eq:merging_time}. The merger rate is than 
\begin{equation*}
    r=\frac{dN(f)}{dt(f)}=\frac{\lambda(f)df}{t_{m}(f)-t_{m}(f+df)}
\end{equation*}
It follows that
\begin{equation}
\label{eq:densità_frequenza}
    \lambda(f)=r\frac{t_{m}(f)-t_{m}(f+df)}{df}=-r\frac{dt_m(f)}{df}
\end{equation}
The spectral density is obtainable by combining eq. \ref{eq:merging_time} and \ref{eq:densità_frequenza}, obtaining a power-law with spectral index $-11/3$:
\begin{equation}
\label{eq:densita_frequenza_espl}
    \lambda(f)=\frac{8}{3}r\frac{f^{-\frac{11}{3}}}{\Delta(\mathcal{M})}
\end{equation}

We infer a total number of super-Chandrasekhar DWD equal to $(2.0 \pm 0.5) \cdot 10^4$ for the frequency band between 1\,mHz and 5\,mHz, $228 \pm 62$ DWDs between 5\,mHz and 1\,dHz and $42 \pm 11$ between 1\,dHz and 5\,dHz. The reported errors are calculated propagating both the error on the rate estimation and the statistical dispersion of the function $\Delta(\mathcal{M})$ (defined in eq. \ref{eq:GW_frequency}). This dispersion is computed from the Galactic DWD population obtained in Sect.~\ref{subsec:SFH}, considering only super-Chandrasekhar systems. As the errors are calculated from a mixed approach, they are intended to be indicative.

Note that the influence of other effects at the merger time does not affect the estimations at lower frequencies, since it can simply be modeled as a time shift in eq. \ref{eq:merging_time}, as long as the main effect that shrinks the orbit during the majority of the inspiral is the GW energy loss.

\subsection{Primordial population and evolution}
\label{subsec:SeBa_and_convolution}
Being the DWD parameter distribution poorly known, the present stellar population can be generated evolving a primordial population with the code \textsc{SeBa}\footnote{\url{https://github.com/amusecode/SeBa}}. This code makes it possible to completely simulate the evolution of a stellar population; in particular, for the sake of this analysis, it is used to evolve a population of binaries in the parameter ranges where the formation of a DWD is possible. We sample the population parameters, namely the masses $m_1$ and $m_2$, the initial orbital separation, the eccentricity and the metallicity $Z$ accordingly with some distributions implemented in the code. The fiducial parameter distributions and parameter intervals are chosen accordingly to \cite{Korol:2017qcx} and reported in tab. \ref{tab:DWD_parameters}. Despite effectively using a thermal distribution for the eccentricity, all the short-period DWDs show $\varepsilon=0$ at the formation of the DWD, at the end of the second CE phase. Thus the effects of eccentricity will be neglected, considering only circular orbits, as anticipated in Sect.~\ref{sec:binary_system}. The CE treatment is implemented in \textsc{SeBa} following the $\gamma\alpha$-prescription, which is thoroughly motivated and discussed in \cite{SeBa_2}. This prescription reproduces well the observations specifically in the context of DD SNIa progenitors, which is the main target population of this paper. After the sampling, every binary system in the population is evolved independently for a chosen time interval, taking into account all the major effects that contribute to the evolution of the single stars and the system as a whole. The state of the system for every relevant evolutionary phase is stored.

\begin{table}
\centering
\begin{tabular}{@{}lll@{}}
\toprule
parameter            & distribution     & interval                     \\ \midrule
Primary mass         & Kroupa IMF       & $0.95 M_\odot<M<10\, M_\odot$  \\
Secondary mass ratio & Uniform          & $0<\frac{M_2}{M_1}\leq 1$    \\
Orbital separation   & Log-uniform      & $1 R_\odot<d<10^{6}\, R_\odot$ \\
Eccentricity         & Thermal          & $0<\varepsilon\leq 1$        \\
Metallicity          &  & $Z = 0.014$                        \\ \bottomrule
\end{tabular}
\caption{Initial population parameters}
\label{tab:DWD_parameters}
\end{table}

The resulting population consists of a sample of binary systems, each with its evolutionary history over a 13.5 Gyr evolution time interval. This corresponds to a ``$\delta$-burst'', from which we obtain the simulated populations through a convolution with the respective star formation histories (SFH).

For every $t$, $0<t<13.5$\,Gyr $\equiv t_0$, in the population a certain number of new systems are formed (systems/year); it is thus possible to define a formation rate $R_{\text{SFH}}(t) \in [0,1]$ by normalizing the SFH of the population at the value that it reached at the peak of star formation. For the time interval $(t, t+ \delta t)$ a fraction $R_{\text{SFH}}(t)$ of the whole \textsc{SeBa} population is sampled and the frequency is evaluated at the present time, namely at the time $\tilde{t}=t_0-t + \xi \cdot \delta t$ in the simulation. The parameter $\xi$ is drawn randomly from the interval $(-0.5, 0.5)$ for every system to account for a uniform distribution within the minimum period $\delta t$, and thus smoothing the $\delta$ functions into a continuous distribution. If the algorithm finds a system that for $\tilde{t}$ lies in the LGWA frequency band, the system is added to the convolved present population. In addition, let $M_s$ be the ``total simulation mass'', namely the sum of the masses of the stars at the time of formation ($\tilde{t}=0$); this characteristic mass is used in Sect.~\ref{subsec:SFH} to correctly account for the various stellar populations within the Milky Way (MW). \footnote{This is the most reliable estimation of the convolved population mass, even if it does not represent the effective mass at the end of the evolution. Obtaining this second estimation would require the complete simulation of the entire population evolution, which is not reasonable considered the necessary computational cost.} Further details and motivations for the use of this procedure are discussed in Appendix \ref{appendix:convolution}.

\subsection{Star Formation Histories}
\label{subsec:SFH}

Once evolved the $\delta$ population sample, the final population is generated via the convolution with a chosen SFH. The spatial distribution of the DWDs (see Sect.~\ref{subsec:spatial_dist}) needs to be performed considering distinct components of the total Galactic population. We consider a central bulge, a thin disk and a thick
disk as described in Sect.~\ref{subsec:spatial_dist}. In principle, also other stellar populations could be simulated, as the Magellanic clouds, the single globular clusters, and finally other galaxies within the horizon of LGWA, to completely model its detection capabilities. However, for this analysis an accurate SFH model is developed for the MW and than reused to model also the extragalactic DWD population.

\begin{figure}
    \centering
    \includegraphics[width=1\linewidth]{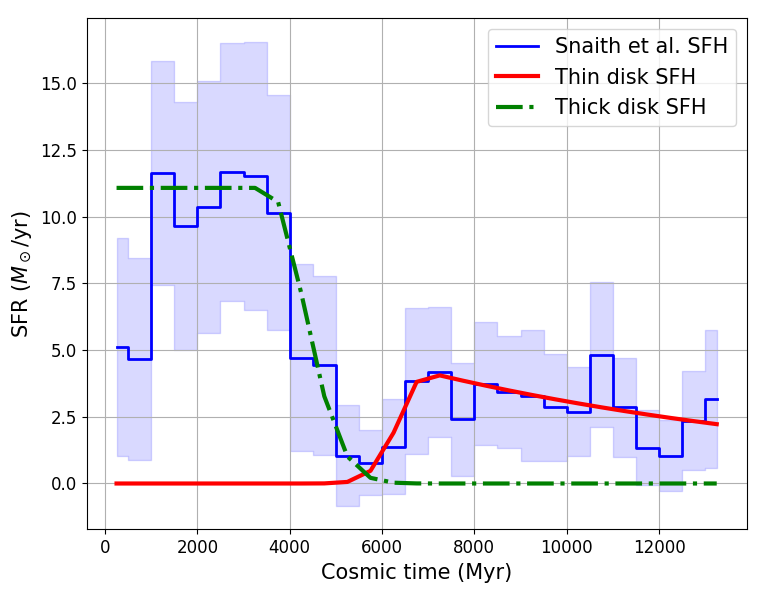}
    \caption{Star formation history of the two Galaxy disks, with data from \protect\cite{SFH_disks} for comparison. The normalization of the SFH is not used for the convolution procedure, but the ratio between normalizations is a byproduct of the entire simulation of the DWD population.}
    \label{fig:F2}
\end{figure}

The SFH of the various components is usually derived from the chemical analysis of the stellar populations; this results in a series of star formation rates at any time from formation up to present days. To simulate the SFH it is chosen a simple analytical approximant that closely resembles the experimental data. The discrepancy between the approximant and the true SFH is negligible, as the global properties of the convolved population are (a posteriori) very robust. The important characteristics of the SFH are the overall abundance and the approximate distribution in time.
The SFH of the disks is chosen in accordance with \cite{SFH_disks, SFH_disks_1}, and presents an early phase of bursty star formation in the thick disk, followed by a clear gap approximately 7.5\,Gyr ago, and a less intense phase in which the thin disk is formed. 
Figure ~\ref{fig:F2} shows the original data from \cite{SFH_disks} with the two approximants for the disks superimposed.

The SFH of the bulge is chosen accordingly to \citealt{Bulge_SFH_TRUE} (based on CCSNe yields by \citealt{Kobayashi_2006}) to be a Gaussian burst with  $\mu=1$\,Gyr, $\sigma=0.66$\,Gyr, followed by a low-intensity exponential tail $\displaystyle\text{SFH}_\text{late}=0.1 \exp{\Big(-\frac{t}{8.3\,\text{Gyr}}}\Big)$ (normalized with respect to the peak of the Gaussian burst). Note that the tail, being nearly quiescent, is less important from the point of view of the total stellar mass produced in the overall SFH, but it is relevant for the final number of present-day DWDs: the binary systems that formed from stars originated in the burst are more likely to have already coalesced. In fact, the fraction of the DWDs in the bulge that were formed in the exponential tail account for the 40\% of the bulge DWD population. A complete consensus on the SFH of the various components is not yet established: for example, regarding the bulge, an opposite model is presented by \cite{bulge_SFH}, with a more intense SFR in the late phase. In principle, with the straightforward procedure described here it is simple to compare different possible SFHs.

The DWD relative abundance among the components is chosen comparing the mass of the simulations $M_{s, \text{bulge}}$, $M_{s, \text{thin}}$, $M_{s, \text{thick}}$ with the estimated masses reported in tab. \ref{tab:MW_parameter} and accounting for the total number of super-Chandrasekhar DWD, $N_{tot}=2\cdot 10^4$ after the estimations presented in Sect.~\ref{subsec:pop_size_estim}. 

Although the normalization of the overall SFH for the three components has been chosen a posteriori via the total mass and the abundance of super-Chandrasekhar DWDs, the ratio between the sampling coefficients of thin and thick disks matches perfectly (better than $1\tcperthousand$) the ratio between the peaks of the SFHs as given by \cite{SFH_disks}, also reported in fig.~\ref{fig:F2}. This demonstrates the robustness of the procedure and the validity of the chosen SFHs. For the bulge, the direct comparison with the original SFH provided in \cite{Bulge_SFH_TRUE} is more difficult, since the conversion from surface SFR density and total SFR is dependent on the chosen spatial distribution (whereas the normalized SFH is not, namely the conversion occurs via a simple multiplicative coefficient). However, since also the bulge SFH influences the sampling coefficients of the disks, this is an indication that also the treatment for the bulge is accurate. 

The mass and frequency distributions of the three populations are plotted and commented in fig.~\ref{fig:FB2}. Note that the differences in abundance and distribution are due only to the chosen SFH, since the three populations are obtained starting from the same original $\delta$ burst.

\subsection{Other parameters}
\label{subsec:other_parameters}
To carry out a complete GW analysis with \textsc{GWFish} it is necessary to fix additional source parameters. The system inclination $\cos \theta$, the polarization angle and the phase are sampled uniformly in the respective domains, the components $a_1$ and $a_2$ of the WDs spins along the orbital angular momentum in units of $m_{1,2}^2$ are sampled uniformly between 0 and 0.1.

\subsection{Spatial distribution within MW}
\label{subsec:spatial_dist}

The mass distribution of the MW is still matter of debate, and growing evidences show that the structure of the Galaxy is very complex \citep{MW_complex_structure}. Since an accurate sampling of a realistic structure is difficult to implement, and is not expected to significantly impact the results compared to other adopted approximations, we adopt a simple model, while still correctly accounting for the overall distribution of the stellar population.

The model consists of a central bulge, a thin disk and a thick disk. In the following a cylindrical coordinate system is used, with the origin in the Galactic center of mass, the $z$-axis as the Galaxy's rotational symmetry axis, $R=\sqrt{x^2+y^2}$ the radial distance from the $z$-axis and $\theta$ the angular variable. This model is thoroughly described in \cite{McMillan:2016jtx}; in addition to being accurate, it is easily implementable in a sampler thanks to the analytical form of the stellar densities. The used parameters are listed in tab.  \ref{tab:MW_parameter}.

The bulge is modeled through the distribution:
\begin{equation*}
    \rho_b(r)=\frac{\rho_{0}}{(1+r'/r_0)^\alpha}\exp{\bigg[-(r'/r_\text{cut})^2\bigg]}
\end{equation*}
 with $r'=\sqrt{R^2+(z/q)^2}$, while the disk density is the sum of two parts: the thin disk and the thick disk, each parametrized with:
\begin{equation*}
    \rho_d(R, z)=\frac{\Sigma_0}{2z_d}\exp{\bigg[-\frac{|z|}{z_d}-\frac{R}{R_d}\bigg]}
\end{equation*}
Despite being distinguishable by the chemical composition, this double disk modeling is used exclusively to account for a better spatial distribution.

\begin{table}
\centering
\begin{tabular}{@{}lll@{}}
\toprule
parameter & value                      & meaning                        \\ \midrule
$r_0$     & 0.075 kpc                   &   B radius, power decrease\\
$r_\text{cut}$     & 2.1 kpc   &  B radius, exponential cut \\
$\alpha$     & 1.8 &    power decrease exponent   \\
$q$     & 0.5 &  B $z$ semiaxis oblation   \\ 
$z_{d \text{thin}}$ & 300 pc & scaleheight of the TnD\\
$z_{d \text{thick}}$ & 900 pc & scaleheight of the TkD\\
$R_{d \text{thin}}$ & 2.53 kpc & scalelenght of the TnD\\
$R_{d \text{thick}}$ & 3.38 kpc & scalelenght of the TkD\\
$M_b$ & $8.9 \cdot 10^9 M_\odot$ & B total mass\\
$M_\text{thin}$ & $3.5 \cdot 10^{10} M_\odot$ & TnD total mass\\
$M_\text{thick}$ & $1.0 \cdot 10^{10} M_\odot$ & TkD total mass\\
\bottomrule
\end{tabular}
\caption{Parameters of the model for the MW components from \protect\cite{McMillan:2016jtx}. "B" = bulge, "TnD" = thin disk, "TkD" = thick disk.}
\label{tab:MW_parameter}
\end{table}

\subsection{Extragalactic sources: spatial distribution and DWD populations}
\label{subsec:extra_rates}
While for the MW the populations are distributed following a multi-component spatial model, for the extragalactic sources this is superfluous: as the distance of a galaxy grows larger than its size, the spatial distribution of the population becomes negligible in determining the detectability and PE of the single DWD. Thus, these sources are associated directly with the position of the respective galaxy centers. For each galaxy within 30\,Mpc (the estimated horizon of LGWA as described in Appendix \ref{sec:plain_analysis}) we extract the following data from the HyperLeda\footnote{http://leda.univ-lyon1.fr/} catalog \citep{Makarov:2014txa}:
\begin{itemize}
    \item Position: $ra$, $dec$ and distance modulus $m_\text{best}$, which is related to the luminosity distance $d_l$ through the following relation\footnote{The parameter $m_\text{best}$ is obtained as a weighted average between two different measurements of the distance modulus: $m_z$, which is measured using the redshift and is more important at higher distances, and $m_0$ which is obtained with independent methods, and is more reliable on small distances.}:
    \begin{equation*}
        m_\text{best}=5\log(d_l)+25
    \end{equation*}
    
    \item $B$ band magnitudes: the apparent magnitude $m_B$ in the $B$ band corrected for Galactic extinction, and the absolute $B$ magnitude $M_B$. 
    \item $K$ band magnitude: the apparent magnitude $m_K$ is used to compute the $B - K$ color. Since the $K$ band refers to the near-infrared spectrum, which is nearly not affected by the dusts, an extinction correction is not applied.
    \item Morphological type of the galaxy, expressed in the Hubble morphological classification.
\end{itemize}

The SN Ia rate for each galaxy in our sample is estimated based on observed SN Ia rates in the local Universe. These rates are provided as functions of various galaxy properties, including luminosity in the $B$ and $K$ bands, stellar mass, and morphological type. Specifically, we adopted the rate–size relations described in \cite{SNIa_rate_Li}, and we used two parameters: the $B$ band luminosity and one chosen among the morphological type and the $B - K$ color. The rates were calculated preferably from the $B - K$ color (obtained as $m_B-m_K$) and the $B$ band luminosity. Where this was not possible due to the lacking of the $K$ apparent magnitude, which is the case for a non-negligible fraction of the population, the morphological type and the $B$ band luminosity were used. See App. \ref{app:extra_rates} for further details. The $B$ luminosity is $L_B=10^{-0.4(M_B-M_{B,\odot})}L_\odot$, where $M_B$ is the absolute corrected $B$ band magnitude and $M_{B,\odot}=5.31 \pm 0.03$ mag is the absolute magnitude of the Sun in the $B$ band. The $B$ luminosity is than expressed in units of $10^{10}L_\odot$ as a reference scale unit for the luminosity of galaxies.

Galaxies lacking both data sets are a minor component (mainly negligible stellar aggregates) and are not expected to significantly affect the overall SN Ia rate. Therefore, they are excluded from the analysis.

\begin{figure}
    \centering
    \includegraphics[width=1\linewidth]{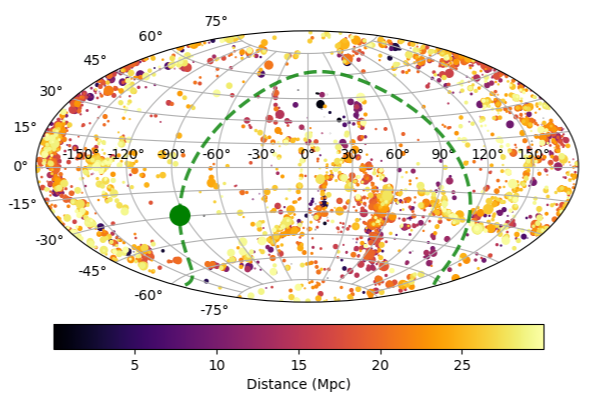}
    \caption{Extragalactic population: the plot shows the position of the galaxies from the HyperLeda catalog; the colorbar indicates the luminosity distance, and the size of every dot represents the rate associated with the relative galaxy. For graphical clarity, the size $s$ of the markers are related to the SN Ia rate $r$ through the relation $s\propto \log(r+1)$. Note the incompleteness of the population along the Galactic plane (green line, the green dot represents the Galactic center).}
    \label{fig:F3}
\end{figure}

The rate-size relations that relate the B band luminosity and galaxy color or morphological type to the rate $r$ are phenomenologically parametrized by \citep{SNIa_rate_Li}:
\begin{equation*}
     r(L_B)=\text{SNuB}(L_0)\cdot L_B\cdot\bigg(\frac{L_B}{L_{B0}}\bigg)^{\text{RSS}_B}
\end{equation*}
Where SNuB is the rate for a galaxy with luminosity $L=10^{10}L_\odot$, and RSS$_B$ is a power-like correction to the simple linear relation between the size (luminosity) and the rate, gauged with respect to a ``typical galaxy'' with luminosity $L_{B0}=2.2 \cdot 10^{10}\,L_\odot$. Both the coefficients depend also on the choice of either color or morphological type as the secondary parameter. The average-size SNuB coefficients and the relative RSS$_B$ from \cite{SNIa_rate_Li} are listed \footnote{Note that since in the HyperLeda catalog $H_0 = 70\, \text{km s}^{-1}\text{Mpc}^{-1}$ is assumed, while in \cite{SNIa_rate_Li} $H_0 = 73\, \text{km s}^{-1}\text{Mpc}^{-1}$, the rates have been corrected by a factor $q = \Big(\frac{H_{0, \text{HyperLeda}}}{H_{0, \text{Li}}}\Big)^2\simeq 0.919$ via 
\[
\text{SNuB}_\text{corr}=\text{SNuB}_{\text{Li}}\cdot q
\hspace{0.7cm} \text{and} \hspace{0.7cm}
L_{B0, \text{corr}} = L_{B0, \text{Li}}\cdot q^{-1}
\]
}
in tab. \ref{tab:SNuB_RSS}. The rates are expressed in units of one SN Ia every 100 years.  A plot of the positions of the galaxies with the corresponding rates and distances is shown in fig.~\ref{fig:F3}. 

The estimation of cumulative SN Ia rate within 10\,Mpc is $31 \pm 7$ SN/100 yrs, $245 \pm 67$ SN/100 yrs for a 20\,Mpc horizon and $650 \pm 157$ SN/100 yrs for a 30\,Mpc horizon. The errors are computed via a Monte Carlo error estimation starting from the errors on the parameters given by \cite{SNIa_rate_Li}, and a 20\% correction has been applied to account for the Zone of Avoidance (see Sect.~\ref{subsec:errori_popolazione}).

\begin{table}
\centering
\begin{tabular}{@{}llllll@{}}
\toprule
type & SNuB($L_0$) & RSS$_B$ & $B - K$             & SNuB($L_0$) & RSS$_B$ \\ \midrule
E           & 0.305       & -0.23   & \textless{}2.3    & 0.158       & -0.25   \\
S0          & 0.282       & -0.23   & 2.3 - 2.8         & 0.152       & -0.25   \\
Sab         & 0.271       & -0.23   & 2.8 - 3.1         & 0.231       & -0.25   \\
Sb          & 0.217       & -0.23   & 3.1 - 3.4         & 0.248       & -0.25   \\
Sbc         & 0.198       & -0.23   & 3.4 - 3.7         & 0.260       & -0.25   \\
Sc          & 0.200       & -0.23   & 3.7 - 4.0         & 0.250       & -0.25   \\
Scd         & 0.165       & -0.23   & \textgreater{}4.0 & 0.305       & -0.25   \\
Irr         & 0.000       & -0.23   &                   &             &         \\ \bottomrule
\end{tabular}
\caption{SNuB coefficients for the rate calculation, from \cite{SNIa_rate_Li} and corrected for cosmology. The reference $B$ band luminosity is $2.2 \cdot 10^{10} L_\odot$.}
\label{tab:SNuB_RSS}
\end{table}

Since the detectable population outside our Galaxy consists only of the most massive DWDs that experience the merging (see Sect.~\ref{sec:results}), only super-Chandrasekhar merging systems are considered.
The realistic number of mergers expected is not sufficient to produce a significant statistics, so a higher number is considered and the results are to be intended as probabilistic, once normalized to the true number of expected events. A broader population is obtained by considering all the merging events produced by the synthetic population in a time span from the nominal observation time $t_0=13.5$\,Gyr to $t_0+2.4$\,Myr. The time interval $\Delta t=2.4$\,Myr is chosen observing that the merger rate of the population is stable for 2.4\,Myr and than drops, due to the fact that the population is depleted from the DWD that merged in the meantime. The merger rate is determined by evolving the separation between the binaries only under GW energy dissipation and considering a Roche scenario as the merging condition. From the resulting population we choose to sample a merging population 500 times larger than the expected one (without incompleteness correction, see Sect.~\ref{subsec:errori_popolazione}).

\subsection{Error budget}
\label{subsec:errori_popolazione}
The produced population can suffer from some errors: the most important will now be listed, with a correspondent estimation of the relative error induced on the synthetic population.
\begin{itemize}
    \item Statistical sampling: given the high number of sampled parameters and the relatively small number of systems in the overall population, the population is subject to considerable fluctuations if the sampling is repeated. This behavior can be modeled with a Poisson statistic.
    \item Initial conditions and parametrization: the population is generated starting from a set of fiducial distributions and parameters. Note however that the total abundance of objects is set independently by the SN Ia rate: for this reason, changes of the initial parameters do not propagate on the global number of super-Chandrasekhar sources, which represent the main target population for LGWA (see Sect.~\ref{sec:results}). The effect of slightly different parametrizations is only a subdominant rearrangement of the mass distribution. We estimate the error contribution from the initial conditions parametrization to the mass and S/N distributions to be $\lesssim3\%$ for the super-Chandrasekhar and $\lesssim7\%$ for the sub-Chandrasekhar  subpopulations. Such effects are analyzed and discussed separately in App. \ref{appendix:initial_conditions}.
    \item Incompleteness of the HyperLeda catalog: as shown in fig.~\ref{fig:F3}, it is not possible to identify the galaxies that are behind the Galactic plane; this will cause a systematic underestimation of the sources that will be visible to LGWA even if obscured by the MW disk for electromagnetic observations. This systematic error does not exceed the $25\%$ of the total population, correspondent to the Zone of Avoidance caused by the obstruction of the MW \citep{Kraan-Korteweg:2000dzx}. The bias due to incompleteness would be between $20\%$ and $25\%$ if not properly corrected. Although we always apply a correction, a potential systematic error of $\approx5\%$ is also taken into account.
    \item Errors in the rate estimation due to uncertainty in the parameters reported in tab. \ref{tab:SNuB_RSS}: the SNuB coefficients present a moderate errors of approximately $5\%$, the corrective parameter RSS$_B$ around $60\% - 90\%$. This is due to the limited amount of SN Ia used to calibrate the parameters. The induced statistical error is $\approx16\%$, obtained by Monte Carlo method in a 30\,Mpc radius. We estimate the potential systematic error to be $\approx 5\%$ by comparing the two possible rate calculation procedures (rates obtained preferably from either color or morphological type, see fig.~\ref{fig:F4}). Correspondingly, the Galactic population inherits from the adopted SN Ia rate an uncertainty of $\approx 20\%$.
    \item Errors in the HyperLeda data: generally negligible when compared with the previous sources of error; the major error source is the estimation of the distance and the color, which can be affected by the extinction. Since a correction for the extinctions is already given in the catalog, this is not considered as a  significant systematic source of error. The morphological type is used marginally in order to recover a minor fraction of the rate, thus errors regarding the morphological type are not relevant.
    
\end{itemize}

The statistical error associated to the results is thus given by the Poissonian statistics; the other sources of error are systematic, since they arise from possible inaccuracies of the global parameters that describe the entire population. The systematic error for the Galactic population is\footnote{ $\oplus$ represents quadratic summation} $20\% \, \oplus\, 7\% \cong 21\%$ (from rate uncertainty and initial conditions for a population including sub-Chandrasekhar binaries), and $16\%\,\oplus\, 3\% \,\oplus\, 5\% = 17\%$ for the extragalactic population (from rate uncertainty, initial conditions for super-Chandrasekhar binaries and potential residual from Zone of Avoidance correction). We remark that these margins are widely conservative in order to avoid a possible underestimation of the errors. In addition, we prefer to report separately the Poissonian and the systematic errors.

\section{Gravitational-wave analysis}
\label{sec:methods}

The analysis of the gravitational-wave signal emitted by the populations generated in the last section is performed by the codes \textsc{GWFish} \citep{Dupletsa:2022scg}, used as a preferred tool to characterize an accurate response, and \textsc{LEGWORK} \citep{Wagg:2022szt, Wagg:2021sgn}, used as an alternative to estimate the S/N in the case that the system is not suitable for the \textsc{GWFish} elaboration. We distinguish between ``inspiraling'' sources, for which the frequency cutoff is due to the end of the observational period, and ``merging'' sources, for which the frequency cutoff is due to the merging of the binary. In particular we provide an analysis of the entire MW population, which consists of inspiraling sources, in Sect.~\ref{subsec:rates_MW}; the MW merging population is treated in Sect.~\ref{subsec:MW_SuperChandra}. Regarding the extragalactic population, in Sect.~\ref{subsec:EXTRA_pop} is presented the analysis of the only super-Chandrasekhar merging population, since as demonstrated in Appendix \ref{sec:plain_analysis} these are the only observable systems outside the MW. In the following we present an introduction to the used codes.

\subsection{\textsc{GWFish}}
\label{subsec:GWFish}
This tool enables accurate simulations of the response of a GW detector to a signal characterized by the parameters simulated in Sect.~\ref{sec:population}, providing an estimation of the errors on the inferred values of the parameters using the Fisher-matrix formalism (see App. \ref{appendix:Fisher}). At present, the \textsc{GWFish} code can provide excellent simulations of the detector's response when the source is relatively close to the merging, but cannot simulate a nearly stationary low-mass source far from the merger. This does not imply that LGWA cannot detect these systems, which can become visible after 10 years of signal integration. For this reason, in the following is presented an analysis of the LGWA sensitivity obtained by \textsc{GWFish} referred only to the merging events. Thus, the \texttt{max\_frequency\_cutoff} parameter is set to be the Roche overflow frequency that characterize the merging (black marks in fig.~\ref{fig:FB2}) for a conservative estimation, or the contact frequency for a more realistic but possibly slightly over-optimistic scenario. Note that during the observational period the probability of observing a Galactic SN Ia is negligible, while the expected SN Ia rate for extragalactic population is such that some merging events are expected within the mission lifetime. \textsc{GWFish} is thus used to precisely characterize merging extragalactic systems and merging super-Chandrasekhar systems.

\subsection{\textsc{LEGWORK}}

We compute an estimation of the S/N for monochromatic MW binaries with the code \textsc{LEGWORK}, originally developed for LISA and thus already implementing the S/N calculation for monochromatic sources.
The LGWA noise PSD provided in the \textsc{GWFish} resources is used to implement the LGWA response into the code. Note that the S/N calculated in this way is only a rough approximation, since the Doppler shift due to the detector proper motion, the source position in the sky and the inclination are not considered for every specific object. The result is a mean S/N, averaged over this parameter subspace, that is anyway useful to estimate global properties of the population.

\section{Results}
\label{sec:results}

\subsection{S/N of the MW inspiraling population}
\label{subsec:rates_MW}
\begin{figure}
    \centering
    \includegraphics[width=1\linewidth]{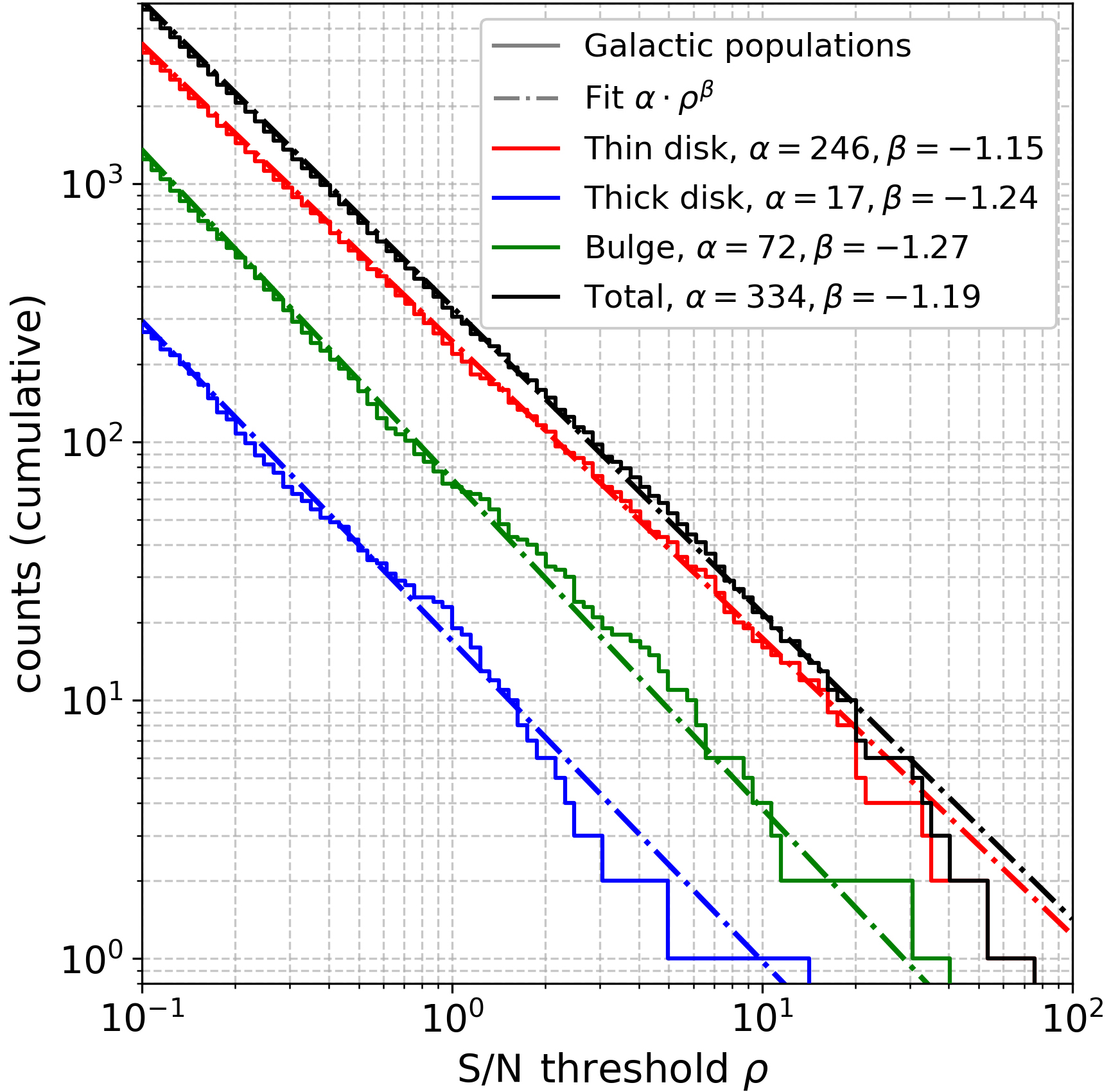}
    \caption{Cumulative distribution of the S/N for the three MW components (colored) and total (black) over ten years of LGWA observation. The plot indicates the cumulative distribution, namely for each S/N threshold $\rho$ (x-axis) the counts (y-axis) provide the  number of objects with the S/N larger than the threshold $\rho$. The power-law fits are also reported explicitly.}
    \label{fig:F5}
\end{figure}

The S/N distribution resulting from the \textsc{LEGWORK} elaboration of the entire (super- and sub- Chandrasekhar) MW inspiraling population is presented in fig.~\ref{fig:F5}. The plot presents the cumulative distribution of the sources: for every $\rho$, the graph gives the number of sources with S/N > $\rho$. This directly quantifies the number of observable objects, once a certain S/N threshold is fixed. This distribution follows a power-law $\Gamma = \alpha \cdot \rho^{\beta}$: the exponents $\beta$ for the three populations are not exactly the same due to different characteristics (mass and spatial distributions), but they are sufficiently similar to allow to fit also the total distribution with a power-law. In addition, the majority of the observable sources comes from the thin disk only. The $\alpha$ coefficients correspond to the number of sources with S/N > 1.  For typical S/N thresholds of 8 and 12, over an observing period of ten years $30\pm 5\text{(stat)}\pm6\text{(sys)}$ and $17\pm 4 \text{(stat)}\pm 4\text{(sys)}$ visible sources are expected, respectively. 

The Galactic inspiraling population emits in the  most sensitive band of LISA, and thus it will not be a prerogative of LGWA only. Repeating the same estimation using the LISA PSD results in a systematically larger S/N. The simultaneous detection of common sources can however be relevant both for scientific and instrumental purposes, such as detector calibration.

\subsection{Super-Chandrasekhar MW merging population}
\label{subsec:MW_SuperChandra}

\begin{figure*}
    \centering
    \includegraphics[width=1\linewidth]{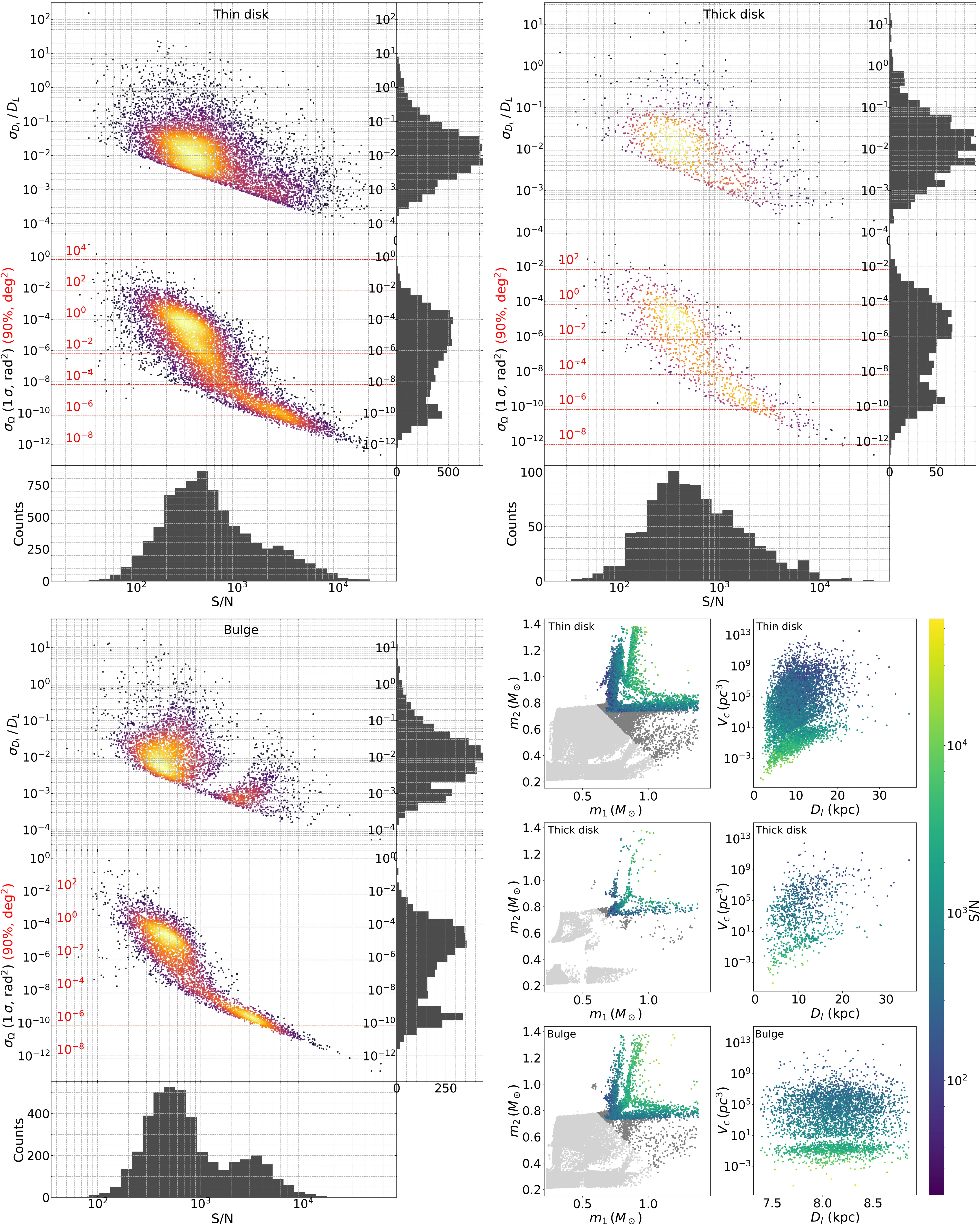}
    \caption{Analysis of the super-Chandrasekhar merging MW population under Roche scenario. The three main plots represent the distribution of S/N, error on sky localization and relative error on luminosity distance for every MW component. Lower left plots: the left column represents the binaries elaborated with GWFish; the colored dots correspond to the elaborated objects with corresponding S/N, gray dots to rejected super-Chandrasekhar binaries and light-gray dots to rejected sub-Chandrasekhar binaries. The right column represents the $1\sigma$ fiducial volume of the elaborated population. The S/N colorscale is common for all the plots.}
    \label{fig:F6}
\end{figure*}

We present a statistical characterization of the MW merging population, although given the SN Ia rate it is unlikely to observe these events.
Not all the simulated systems are treatable by \textsc{GWFish}, as we show in Fig. \ref{fig:F6}, lower right plots: the colored dots represent the elaborated objects with corresponding S/N, the rest are rejected by the code (super-Chandrasekhar systems represented in dark gray). The lacking population corresponds approximately to the fraction of the super-Chandrasekhar population with $m_2<0.73 \, M_\odot$, which has a lower maximum frequency cutoff due to the larger radius of the lighter companion following the mass-radius relation. However, the S/N is still high at the boundary, sign that also this population can be detectable at the merging. The DWDs that can be simulated present a S/N$>10^2$, with a typical S/N$\approx 4\cdot 10^2$ and some well above this typical value.

Fig.~\ref{fig:F6} shows the distribution of the localization parameters (S/N, relative error on luminosity distance and error on angular sky location) within the three MW populations and extragalactic population (see Sect.~\ref{subsec:EXTRA_pop}).
The plots are structured to present the population density (colorscale) in the S/N vs $\sigma_{D_l}/D_l$ space (upper plot) and in the S/N vs $\sigma_\Omega$ space (lower plot); the middle plot have two vertical scales, one in black that represents the $1\sigma$ error (standard deviation) of the sky location, measured in square radians, and a secondary red scale that represents the 90\% sky area in square degrees, as a widely used measure for the sky localization precision. Finally, on the sides the marginal distributions are reported as histograms.

The distribution in the upper plots of every graph (relation between the relative error of the luminosity distance and the S/N) features a neat border of the population distribution. This effect is expected, and explained in Sect. \ref{app:SNR_Dl_limit}. The best angular precision corresponds to $\approx0.3$ arcsec, and the most probable value to $\approx 330$ arcsec, both expressed at CL = 90\%. This makes the LGWA resolution at the level of precise EM observatories within the framework of galactic high-S/N merging events.

Given the errors on sky localization and luminosity distance, it is possible to estimate the volume that contains the source as the portion of space identified by the $1\sigma$ errors on the localization parameters. This is a fundamental parameter, as accurately locating a source requires it to be confined within a reasonably small volume. The typical distances between stars in the MW strongly depends on the position, but an order of magnitude of the typical distances is around $0.1$ to $10$ pc. This means that the volume needed to identify a single star should be $ \lesssim 10^{-3}$ pc$^3$, although for the population near the Solar System, more accessible to optical observations, higher volumes can be enough. In addition, optical characterization of the possible sources (e.g. via spectral classification) can further increase the minimum volume if needed. In fig.~\ref{fig:F6}, bottom right plots, it is presented the relation between the distance and the volume: only a negligible part of the MW population lies under the (very strong) limit of $10^{-3}$ pc$^3$, however the majority the systems are limited in a maximum volume of less than $10^3$ pc$^3$, which is enough to apply other identification techniques to better constrain the position of the source if it is visible to electromagnetic detectors.

\subsection{Extragalactic population}
\label{subsec:EXTRA_pop}

\begin{figure*}
    \centering
    \includegraphics[width=1\linewidth]{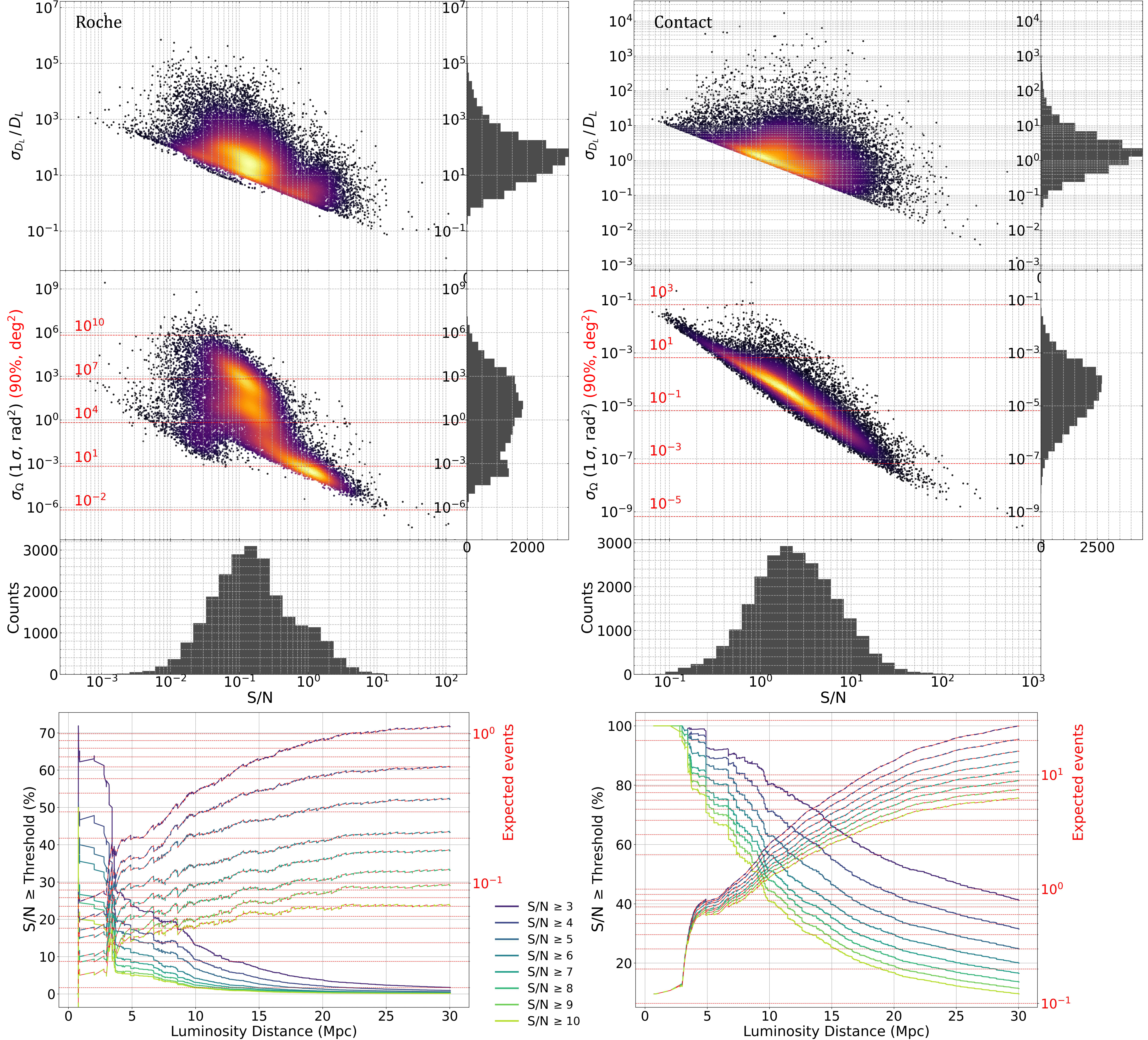}
    \caption{Extragalactic population. Left column: Roche overflow scenario. Right column: contact scenario. Upper row: localization and detection performance. Lower row: cumulative detections (detection percentage and total number of detections) for different S/N thresholds.}
    \label{fig:F7}
\end{figure*}

For the extragalactic population the fraction of the systems that can't be elaborated is $\approx 8\%$ of the population. Given the estimation of the SN Ia rate obtained in Sect.~\ref{subsec:EXTRA_pop} associated with a local rate density of $30.8$ SN Mpc$^{-3}$ (100 y)$^{-1}$, it is a priori expected to observe 1 or 2 SN events in a 10\,Mpc radius during the 10 years observational period.

The detection and localization performances are presented for the two merging scenarios in figure \ref{fig:F7}; the two upper plots are analogous to the ones presented in fig.~\ref{fig:F6} for the Galactic population. It's clear that in the Roche scenario the majority of the events are not detectable; this changes considering the contact scenario, in which also very high S/N are possible. The lower plots represent the relative and absolute abundance of observable sources as a function of the distance, for the two scenarios. This representation is useful in order to normalize the test population (500 times larger than the real) to the actual rates. For each scenario and for a set of 8 S/N thresholds between 3 and 10 are shown two quantities: the colored uniform line, relative to the left vertical scale, shows the percentage of systems that are detectable given the S/N threshold within the reported distance. The dashed lines represent the total absolute number of detections given a S/N thresholds within the reported distance, and are obtained by multiplying the percentage of detected systems by the previously found rates, accounting for the mission duration and the incompleteness factor. In the Roche scenario this value saturates quite early at low values, indicating that observable events with these conditions are not expected. On the contrary, for a ten years observational period the contact scenario predicts a total $10 \pm 3\text{(stat)}\pm 2\text{(sys)}$ detectable events for a S/N threshold of 8, and the total number is not saturated yet at 30\,Mpc, hinting at the possibility to observe events even further. It is thus extremely prominent to better characterize the expected physics of the merging, and in particular the merging frequency, as this is the main parameter that determines the observability of the merging DWD given the operative frequency window of LGWA.

Finally, to estimate a maximum volume $V_\text{max}$ that can be tolerated to correctly discriminate the host galaxy (confusion limit), the HyperLeda catalog is once again used: since the MW is located in a mass overdensity, the confusion limit is estimated in three different "shells": from $D_l=0$ to 10\,Mpc, from 10 to 20\,Mpc and finally from 20 to 30\,Mpc. The average confusion limit is obtained as the inverse of the numerical density of galaxies in the corresponding spatial shell. The resulting confusion limits are 10, 15.5 and 29.8\,Mpc$^3$ respectively.
The $1\sigma$ volumes relative to the detected events, with a S/N threshold of 5, are nearly all under these limits in the respective shells. This means that for merging DWDs the confusion limit is nearly never surpassed, due to the inevitable S/N threshold that limits the detection. Thus LGWA can fulfill the requirements needed to perform this type of identification, under the more general conditions of detection. 
In addition, we provide in tab. \ref{tab:summary} the typical localization capabilities for the sub-population with S/N>8 under the contact scenario hypothesis.

\section{Conclusions}

The LGWA detector will give access to the dHz frequency band in the GW spectrum, which is expected to contain a multitude of observable objects and new physics; in particular the DWDs merge in this band, and are possible candidates for SN Ia progenitors, as discussed in Sect.~\ref{sec:1_DWD_general}. A fiducial synthetic population of DWDs has been generated, following the latest and most reliable physical models regarding stellar structure, star formation, stellar evolution and mass distribution within the Milky Way and for the extragalactic population, calibrating the size of the populations with the expected SN Ia rates (see Tab. \ref{tab:summary}). The population synthesis highlights a crucial feature of the short-period DWD mass distribution, namely the existence of a super-Ch branch: the super-Ch binaries are clustered in a branch in which the most massive star has a mass ranging from $\approx 0.8 \, M_\odot$ to $M_\text{Ch}$ with a  $\approx 0.8 \, M_\odot$ companion. The region above the super-Ch branch in mass space is not populated. This is important when considering the possible masses of a test binary system while evaluating the theoretical performances of a GW detector. In particular, it's more meaningful to compare different systems inside the super-Ch branch (thus with $m_2 \simeq 0.8\,  M_\odot$ fixed and $m_2 \in [0.8, 1.4]\, M_\odot$) than comparing equal-mass systems as it is often done.

Afterwards, an analysis of the detectability of the population binaries was performed using the codes \textsc{GWFish} and \textsc{Legwork}. For Galactic spiraling DWDs we found a power-law S/N cumulative distribution with spectral index $\beta = -1.19$ and $N(\text{S/N}>1)=334$. Considering a representative S/N threshold $\rho=8$, this implies $N(\text{S/N}>8)=30$. For merging DWDs, LGWA can effectively localize and thus identify a significant fraction of the sources as a single systems in the MW and as a single galaxies for the extragalactic population. The precision that should be reached by the detector is generally enough to locate the object unambiguously and link it to a potential electromagnetic observation, under the wider requirement of detection. The confusion limit however can be reached by some objects, implying that other studies are advisable on this topic, covering at least two aspects: 
\begin{itemize}
    \item A more detailed estimation of the properties of the expected sources, in particular for what regards the exact merging frequency of the DWDs, which introduces a frequency cutoff that is crucial to determine the S/N and the localization precision. A small variation of the frequency cutoff implies a considerable change in the S/N and localization quality.
    \item A precise modelization of the waveform, which is heavily affected by mass effects and does not correspond to a pure general relativity waveform, especially in the latter phases of the merging. Since this last phase could account for a significant fraction of the S/N of the event, it's essential to produce a set of reliable approximants.
\end{itemize}

In properly covering the parameter space, this future modelizations should consider with care in particular the systems within the super-Ch branch, for the reasons explained before. Already a coarse-grained sampling of the super-Ch branch would be of great utility to clarify the merging frequency indetermination issue, which represents by far the main uncertainty in the foreseen performances of LGWA.

The detection rates depend on the chosen S/N threshold, and are presented in Sect.~\ref{subsec:rates_MW} for the spiralling binaries inside the MW and in Sect.~\ref{subsec:EXTRA_pop} for the merging systems outside the Galaxy. We provide a summary of the main quantitative findings in Tab. \ref{tab:summary}. The expected detection rate for a S/N threshold of 8 over a ten years observation period are $30\pm 5\text{(stat)}\pm6\text{(sys)}$ spiralling binaries inside the MW and $10\pm 3\text{(stat)}\pm2\text{(sys)}$ extragalactic mergings, under the hypothesis of the contact scenario. Although this estimate is susceptible to fluctuations depending on the true dynamics of the merging process and the true correlation between a super-Chandrasekhar DWD merging and a SN Ia event, which we assumed being exact, our results clearly demonstrate the transformative impact of LGWA on our understanding of DWD populations and their role as Type Ia SN progenitors.

\begin{table*}[]
\begin{tabular}{@{}ll|llll@{}}
\toprule
                                                      &                & \multicolumn{2}{l}{Galactic DWD (spiraling)}                                                               & \multicolumn{2}{l}{Extragalactic DWD (merging)}                                                                                                                                                                               \\ \midrule
quantity                                              &                & value                                  & \multicolumn{1}{l|}{ref. in text}                                 & value                                                   & ref. in text                                                                                                                                                                 \\ \midrule
Detectable  (S/N \textgreater 8) &                & $30\pm 5\text{(stat)}\pm6\text{(sys)}$ & \multicolumn{1}{l|}{Sect. \ref{subsec:rates_MW}} & $10\pm 3\text{(stat)}\pm2\text{(sys)}$                  & Sect.~\ref{subsec:EXTRA_pop}                                                                                                                                \\ \midrule
Total abundance &                & $(5.5\pm 0.007 \text{(stat)}\pm1\text{(sys)}) \cdot 10^5$ & \multicolumn{1}{l|}{} &              &                                                                                                                                \\ \midrule
Super-Ch. abundance &                & $(2\pm0.01\text{(stat)}\pm0.4\text{(sys)}) \cdot 10^4$ & \multicolumn{1}{l|}{Sect. \ref{subsec:pop_size_estim}} & $50 \pm 7 \text{(stat)}\pm9\text{(sys)}$             &                                                                                                                                \\ \midrule
\multirow{2}{*}{$90\%$ sky area [deg$^2$]}             & median         &                                        & \multicolumn{1}{l|}{}                                             & $8.6 \cdot 10^{-3}$                             & \multirow{6}{*}{\begin{tabular}[c]{@{}l@{}}Sect.~\ref{subsec:EXTRA_pop},\\ Fig. \ref{fig:F7}\end{tabular}} \\
                                                      & 95\% interval &                                        & \multicolumn{1}{l|}{}                                             & $\Big[3.6 \cdot 10^{-4}, 1.5\cdot 10^{-1}\Big]$ &                                                                                                                                                                              \\ \cmidrule(r){1-5}
\multirow{2}{*}{$d_L$ relative error}                 & median         &                                        & \multicolumn{1}{l|}{}                                             & $4.9\cdot 10^{-1}$                                      &                                                                                                                                                                              \\
                                                      & 95\% interval &                                        & \multicolumn{1}{l|}{}                                             & $\Big[6.3 \cdot 10^{-2}, 1.5\cdot 10^{1}\Big]$          &                                                                                                                                                                              \\ \cmidrule(r){1-5}
\multirow{2}{*}{Fiducial volume [Mpc$^3$]}                      & median         &                                        & \multicolumn{1}{l|}{}                                             & $1.5\cdot 10^{-3}$                       &                                                                                                                                                                              \\
                                                      & 95\% interval &                                        & \multicolumn{1}{l|}{}                                             & $\Big[2.9 \cdot 10^{-7}, 2.6\cdot 10^{-1}\Big]$ &                                                                                                                                                                              \\ \bottomrule
\end{tabular}
\caption{Summary table of the significant quantitative findings. The merging case refers to the contact scenario, as in the (very conservative) Roche scenario the extragalactic detection rate is negligible. The total and super-Chandrasekhar abundances refer to the systems with $f>1$ mHz and $d_L<30$ Mpc. The observation time is 10 years in both cases.}
\label{tab:summary}
\end{table*}

\begin{acknowledgements}

We wish to thank Enrico Cappellaro for the useful advices regarding SN Ia rates and the anonymous Referee for the valuable revision of the manuscript. We acknowledge the usage of the HyperLeda database (http://leda.univ-lyon1.fr). MB and JH acknowledge the ACME project which has received funding from the European Union's Horizon Europe Research and Innovation programme under Grant Agreement No 101131928.
\end{acknowledgements}


\bibliographystyle{bibtex/aa} 
 \bibliography{biblio.bib} 

\begin{appendix}
    \section{Observability analysis}
\label{sec:plain_analysis}

Before simulating a realistic population, it is useful to consider the observability and parameter estimation of the detector on an averaged population with simple uniform prior distributions. In particular this approach is useful to understand the sensitivity of the detector for the extragalactic DWD merging population. Following the approach of \cite{LGWA_IMBH}, the S/N and the error on the estimation of the luminosity distance are evaluated for a simple DWD population, varying the masses and the luminosity distance. In particular these quantities are evaluated for every triplet $(m_1, m_2, d_L)$ with $m_1, m_2$ in the interval $[0.6\, M_\odot, 1.4\,M_\odot]$ at steps of $0.1\,M_\odot$ and $d_L$ in a set of 6 representative distances of 2, 5, 10, 20, 30, 40\,Mpc. For every combination, a sample of 20 system is analyzed with \textsc{GWFish}, completed with the other parameters ($\cos \theta$, $\alpha$, $\sin \delta$, $\psi$, $\phi$) drawn uniformly in their respective domains. Note that, since we are interested in merging signals, a frequency cutoff has been introduced following the Roche overflow merging hypothesis treated in Sect.~\ref{subsec:merging_frequency}, using the explicit analytical formulas provided. From this sample is than extracted the mean of the S/N, the mean error on the luminosity distance, the best S/N and the smaller error on $d_L$. Since LGWA will be meant to study both the overall statistical characteristics of the population and the single high-S/N cases, these quantities are good indicators of the performances in the two cases.

\begin{figure*}[htbp]
    \centering
    \subfloat[The colorscale indicates the S/N. The estimations for the mean S/N is located over the diagonal, whereas the best S/N is reported under the diagonal. The dots with red edge correspond to S/N < 5, and are thus not likely to be observed.\label{fig:FA2a} ]{
        \includegraphics[width=\textwidth]{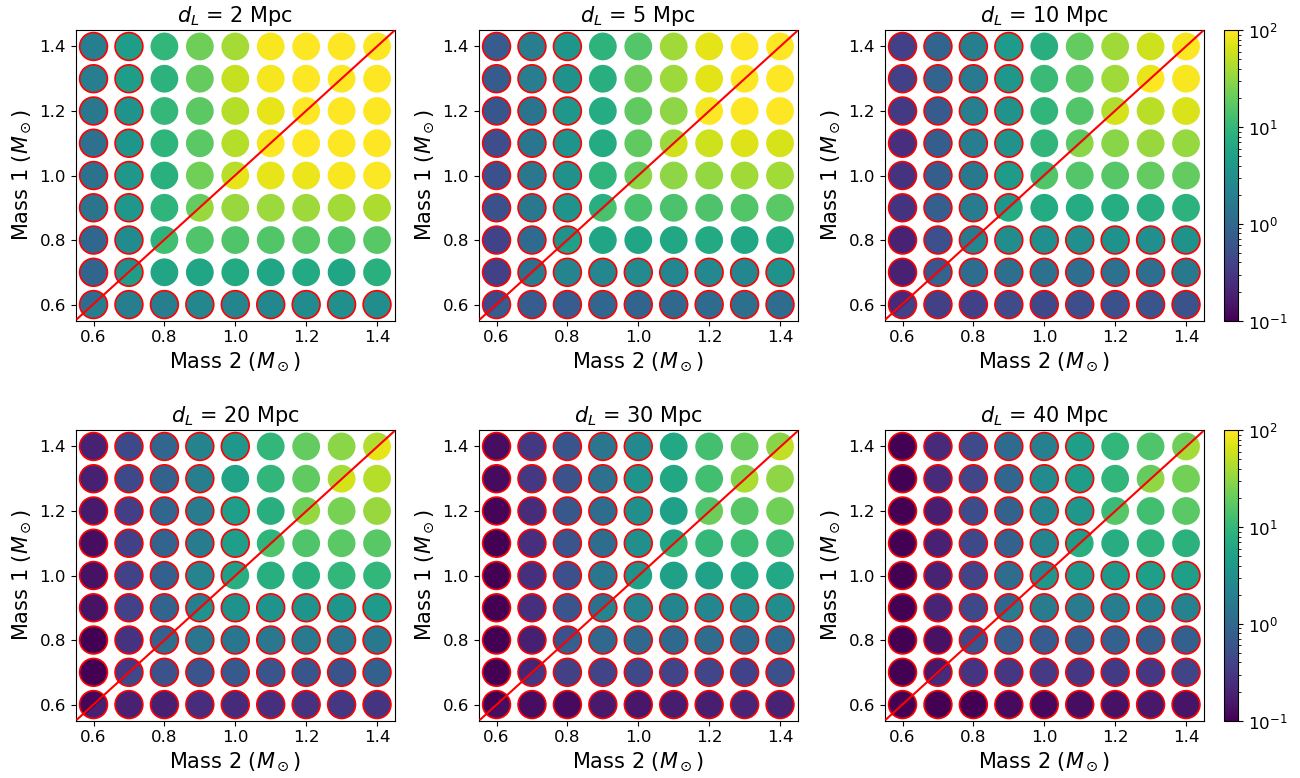}
    } \\
    \subfloat[The colorscale indicates the error on the luminosity distance, in units of Mpc. As in figure \ref{fig:FA2a}, the estimations for the mean error is located over the diagonal, whereas the smaller error is reported under the diagonal, the dots with red edge correspond to S/N < 5. \label{fig:FA2b}]{
        \includegraphics[width=\textwidth]{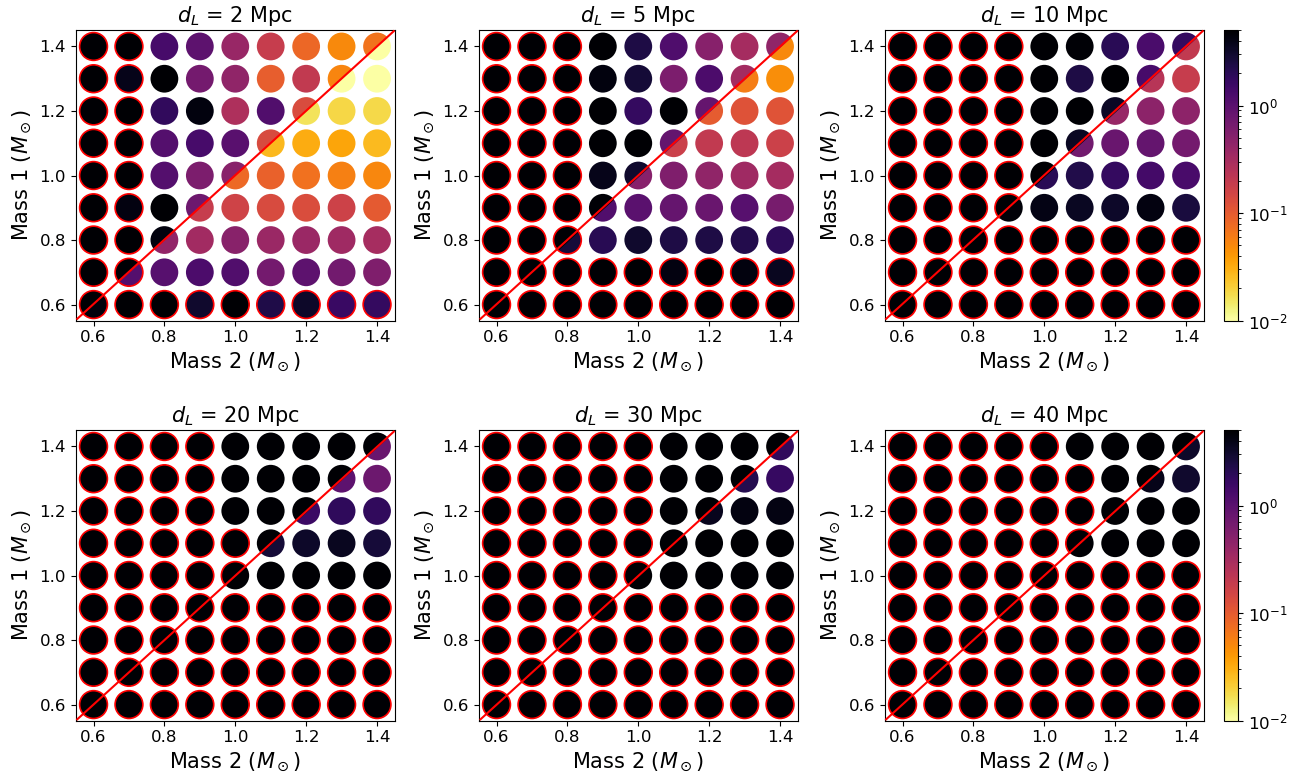}
    }
    \caption{Comparison of the mean and best S/N (upper plot) and best error on luminosity distance estimation (lower plot) for different mass combinations and luminosity distances. The estimation is performed assuming the Roche merging scenario.}
  \label{fig:FA2}
\end{figure*}

The results for the S/N are shown in fig.~\ref{fig:FA2a}. The plot reports over the diagonal the mean S/N, and the best S/N under the diagonal. The S/N is encoded by the colorscale. the red-edged dots correspond to S/N < 5, namely those systems which will be surely not seen because of the S/N threshold. Here a S/N threshold of 5 is chosen as a simple possible representative value. The choice of the mass range is justified by the fact that lower masses will have, even in the best-case scenario, a S/N that does not allow the detection.  It is evident that the S/N is mainly determined by the mass of the lighter WD in the binary, since it is the one that will overflow the Roche lobe, setting a maximum frequency cutoff. As said previously, the most interesting range is the super-Ch branch, which is quite well probed at distances $<5 $\,Mpc and begins to fade with increasing distance.  Note that, for visualization purposes, the maximum S/N shown is 100, but for $d_L = 2$ and 10\,Mpc the S/N of the most massive systems saturates, reaching up to $\approx 10^3$. The main difference between the averaged systems and the better systems is mainly the possibility to access masses of $0.1 \, M_\odot$ less for the best systems (and consequently a global equivalent shift in the S/N value towards higher S/N). Some systems are observable even beyond 40\,Mpc, but it is clear from the synthetic population shown in fig.~\ref{fig:F1} that the systems in this region are really few.

The results for the error on the luminosity distance are plotted in fig.~\ref{fig:FA2b}, with the same convention adopted for fig.~\ref{fig:FA2a}. The colorscale is in units of Mpc. The boundaries of the colorscale have been chosen to have a maximum of 5\,Mpc, since beyond this threshold it becomes impossible to correctly locate the host galaxy. In this case it's evident the strong difference between the average case and the best case for nearby systems. This is due to a large variability of the error, that strongly depends on the choice of the remaining parameters which are randomly drawn. Although it is already evident that a good estimation of the host galaxy can be done only below 10\,Mpc, a more complete treatise of this estimation is presented in section \ref{sec:methods} on the realistic DWD population. These results enforces the fact that the interesting systems lie approximately below 30\,Mpc.

\begin{figure}
    \centering
    \includegraphics[width=1\linewidth]{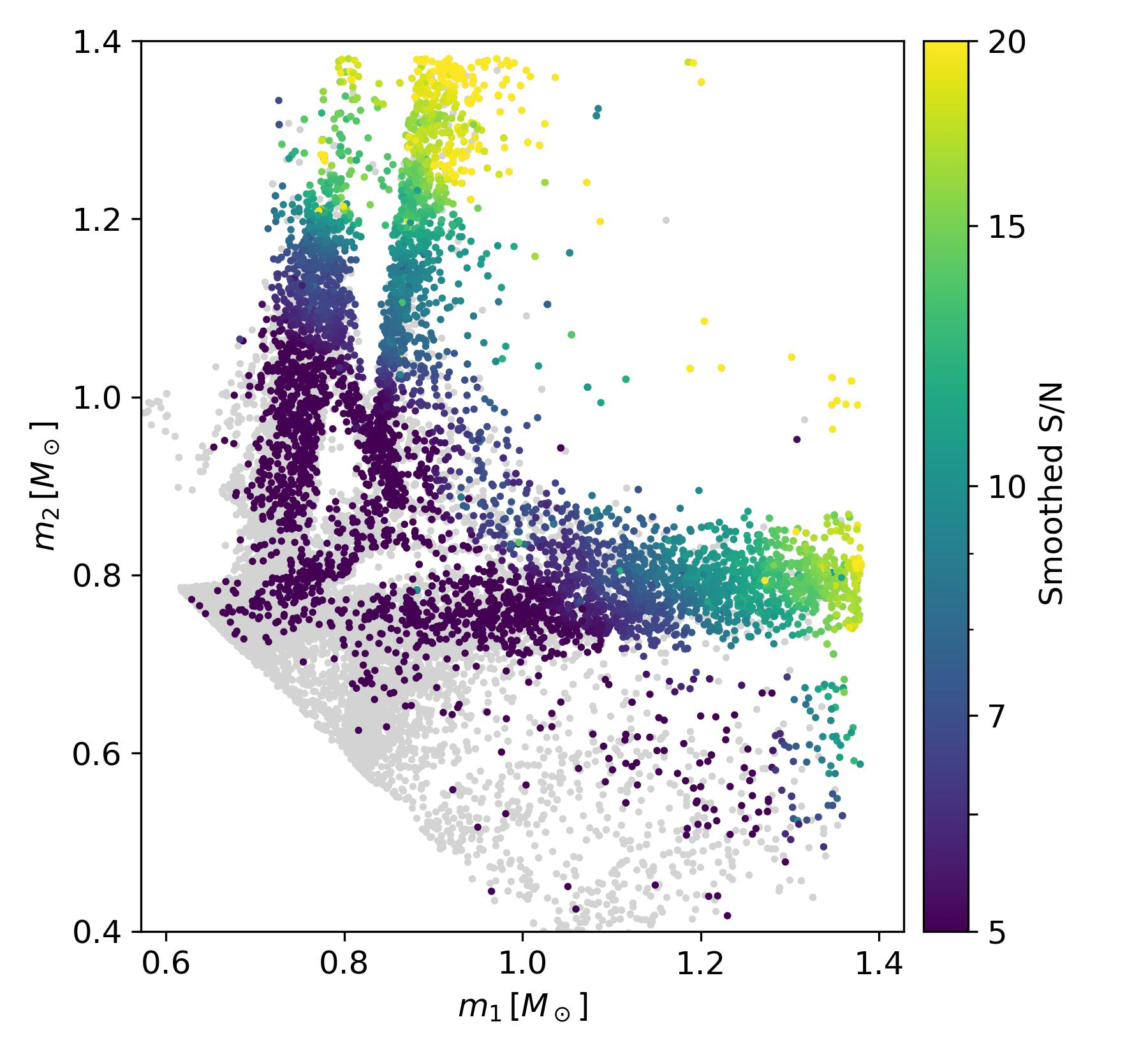}
    \caption{Smoothed S/N (with an averaging radius of 0.05 $M_\odot$) for the extragalactic population under the contact scenario assumption. Only the systems with individual S/N>5 are colored, the gray population presents a lower S/N.}
    \label{fig:FA1}
\end{figure}

As a confirmation of these results, in fig. \ref{fig:FA1} is plotted the mass distribution of the super-Chandrasekhar merging extragalactic population, displaying the S/N for only the systems with S/N>5 for the contact scenario, namely the most optimistic. The displayed S/N is ``smoothed'', namely it's averaged in the mass space over a radius of 0.05 $M_\odot$ in order to show the mean dependence of the S/N from the mass combination. It's evident that only the super-Chandrasekhar systems are detectable outside the Galaxy.

\section{Methods}
\label{appendix:methods}
\subsection{Convolution procedure}
\label{appendix:convolution}
The population resulting from the \textsc{SeBa} sampling and evolution (Sect.~\ref{subsec:SeBa_and_convolution}) consists of $2\cdot 10^{7}$ binary systems; approximately $75\%$ of the systems become DWD at some point of the evolution and among these only $\approx 5.8\%$ at the end of the CE phase display an orbital separation which is sufficiently small in order to enter the mHz frequency band at some time before the end of the 13.5\,Gyr evolutionary period, due to the GW emission. Finally, only few hundreds fall in the LGWA frequency band exactly at the end of the evolutionary period, all around the mHz and none above 10\,mHz. Thus the evolution of a single burst cannot contain a statistically relevant sample of the population, since the expected DWD Galactic population in the mHz should account for $\approx 2\cdot 10^{4}$ systems above $M_\text{Ch}$ in order to reproduce the current SN Ia rate (see the estimations in Sect.~\ref{subsec:pop_size_estim}), and the simulation can't reproduce at all the rare but crucial population with $f_{GW}>0.01$ Hz in the most sensitive band of LGWA.

The solution comes from using the generated population as a $\delta$ function and convolve it with a chosen star formation history (SFH), as detailed in Sect.~\ref{subsec:SeBa_and_convolution}. This approach solves the initial problem as not only the few short-period DWDs that remain at $t=13.5$\,Gyr are considered, but all the systems that at some point of the 13.5\,Gyr evolution had become a DWD. In this way the computationally expensive \textsc{SeBa} simulation is better exploited. In addition, the $\delta$-convolving approach benefits from two main advantages:
\begin{itemize}
    \item A SFH is naturally simulated with accuracy as the convolution of the chosen $R_\text{SFH}(t)$ with a Dirac $\delta$ gives the SFH itself; this  allows one to test multiple SFH hypothesis without running the \textsc{SeBa} simulation multiple times, since only one run is sufficient to generate the $\delta$ burst ($\approx$ 1200 CPU hours have been used). In addition, different Galactic components have different SFHs, hence this approach is ideal to simulate a complex Galactic structure.
    \item A large DWD population can be easily produced, allowing for more precise statistical considerations. This could not be possible with direct evolution simulation, given the extremely low fraction of observable systems in the total population. 
\end{itemize}

However, there are some drawbacks to consider:
\begin{itemize}
    \item The $\delta$ burst refers to fixed initial parameters that could change over time in the real SFH. For example, the metallicity is expected to increase as the Galaxy ages or vary for the different Galactic populations (bulge, thin disk, thick disk, see Sect.~\ref{subsec:spatial_dist}), but this effect cannot be taken into account since a variation in the metallicity would require to re-run the evolutionary simulation. The same is true for the IMF and the binary fraction.
    \item The same DWD system is counted more than once if $\delta t$ is small enough, resulting in a certain degree of correlation inside the population.
\end{itemize}

For the former, there is not a simple solution, and it will contribute for some degree of systematic error. This problem however is strongly mitigated by the choice of the SFH: in general, it is expected that $R_{\text{SFH}}$ in the early stages of Galaxy evolution is far higher than in the late stages (sec.~\ref{subsec:SFH}); this results in a shorter effective formation period, that can more acceptably be modeled with fixed metallicity and IMF function. In particular, the population dependence on the metallicity is investigated systematically in \cite{Korol:2020lpq}, where it is shown that the effect of drastically changing the metallicity (over 3 order of magnitudes) is only a mild variation of the DWD abundance. In our case the abundance is independently gauged by the SN Ia rate, thus this effect is not expected to play a significant role.

For the second point, it is important to estimate how much time a system remains in the LGWA band. An upper limit is found using the expression for the time to merge for a binary system, eq. \ref{eq:merging_time}. The initial frequency for which the algorithm recognizes the DWD as detectable with LGWA is $f_i=1$\,mHz. The resulting merger times for the typical DWD masses is around 1 -- 3\,Myr. This means that the time resolution $\delta t$ should be lower than 1\,Myr, otherwise some systems would be lost in the process, but not too low to cause detectable correlation in the population. Note that lowering $\delta t$ is in principle desirable, as doing so the resulting population would increase in number. A lower limit for $\delta t$ can be found imposing that the frequency space density generated by the repetition of one system should be much lower than the density of different systems. This constrain prevents the formation of isolated "clusters" in the frequency space due to the multiple counting of a single system. With really rough estimations (considering one half of the systems in the interval 1 -- 5\,mHz and imposing there the condition, since the low-frequency band corresponds to a slower evolution) it is found that this limit $\delta t_\text{min}$ always lies under 0.001\,Myr. The real limit must be even lower, since all the entire estimation is conservative, but even this value is far from the $\delta t$ required for the generation of the complete population. \footnote{Note that the bigger is the initial \textsc{SeBa} sample, the higher will be the different-systems density and thus the smaller the minimum $\delta t$, so the maximum potential size of the convolved population grows faster than linearly with regards to the \textsc{SeBa} population size.}. In conclusion, for the chosen initial population size, taking $\delta t \in [0.01, 1.0]$ Myr is a legitimate choice that does not introduce a significant correlation while exploiting at the best the DWD systems that had been previously elaborated. To properly simulate the MW population, it is sufficient a time interval $\delta t = 1$\,Myr.

\subsection{MW populations}
\begin{figure}
    \centering
    \includegraphics[width=1\linewidth]{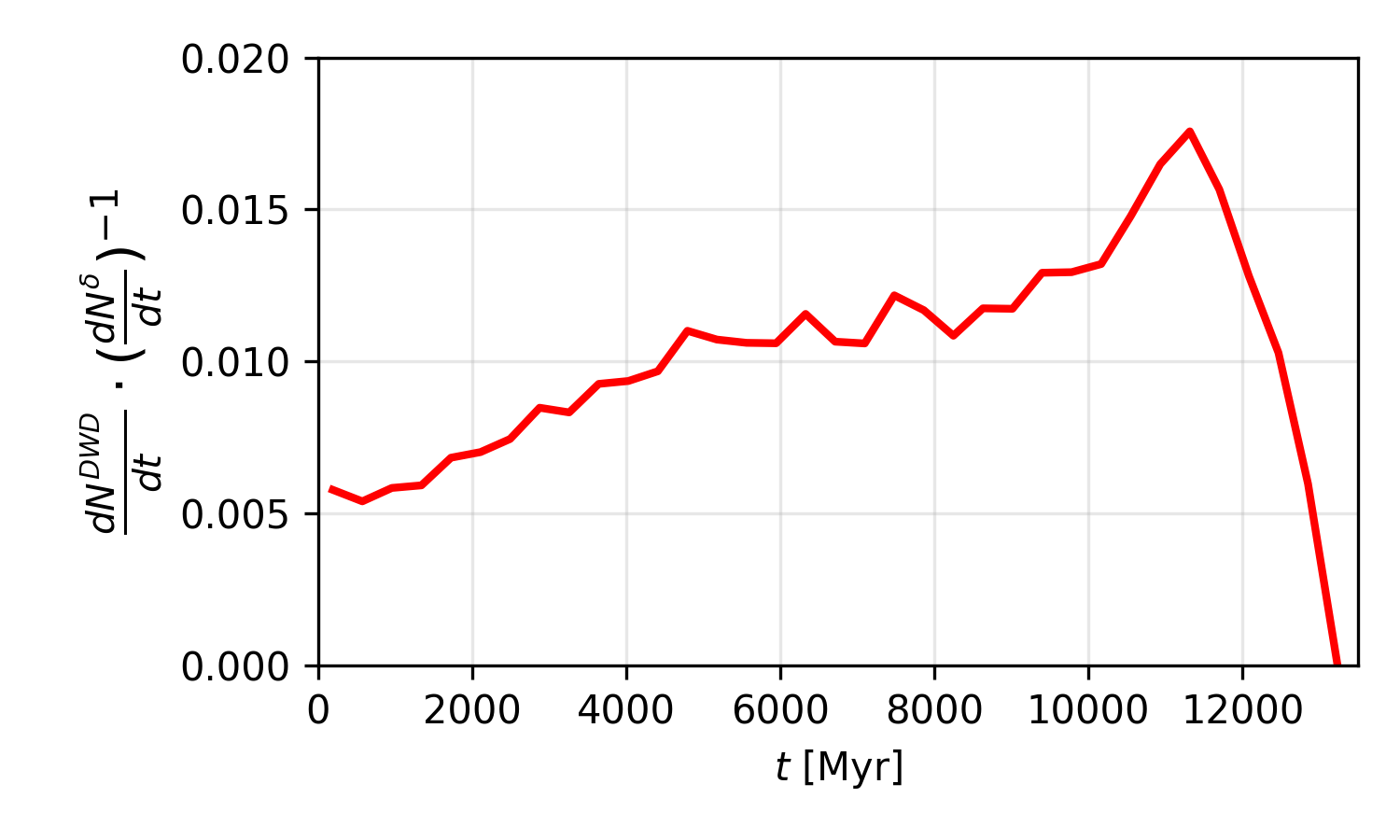}
    \caption{Present-DWD gain density as a function of time.}
    \label{fig:FB1}
\end{figure}

\begin{figure*}
    \centering
    \includegraphics[width=1\linewidth]{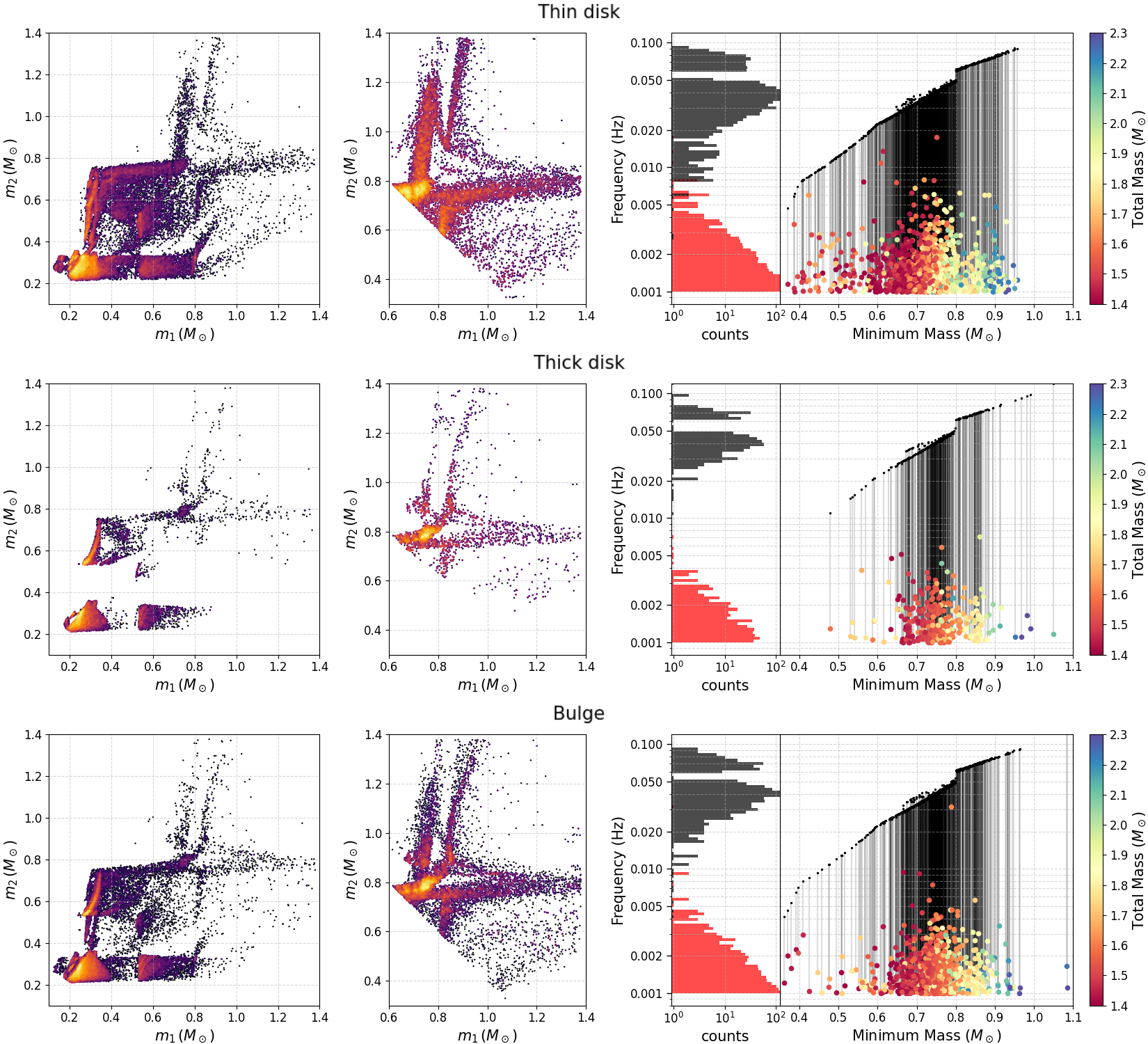}
    \caption{The three simulated population components of the MW (row 1: thin disk, row 2: thick disk, row 3: bulge). Left: mass distribution of the binaries, the color represents the density for more clarity, 10\% of the population is showed. Center: mass distribution of the super-Chandrasekhar subpopulation; all the population is shown. Right: distribution of the super-Chandrasekhar population as a function of the mass of the less massive WD of the system (x-axis), frequency during the LGWA observation period (colored marks, total binary mass as colorscale) and final merging frequency in Roche scenario (black marks). To the left side the marginalized histograms of the population in frequency domain are plotted. The histograms for the frequency distribution during the observing period are shown in red, the ones relative to the final Roche merging frequency in black. 5\% of the Super-Chandrasekhar population is shown.}
    \label{fig:FB2}
\end{figure*}

We visually report the main properties of the three convolved populations plotting the mass and frequency distributions of the sources in fig. \ref{fig:FB2}. As highlighted in Sect.~\ref{subsec:SeBa_and_convolution}, the differences between the populations arise only from the choice of different SFH and the total mass of the respective components. Early or late periods of SFH contribute differently to the number of DWD at present time $t_0$: in order to visualize the different gain in terms of DWD production for different periods we plot in fig. \ref{fig:FB1} the number density of DWD coming from progenitors produced at time $t$ in a time bin $dt$, namely $\displaystyle\frac{dN^\text{DWD}}{dt}$, normalized to the number density of the considered progenitors $\displaystyle\frac{dN^\delta}{dt}$ for the same value of $t$. The plot effectively represents the fraction of systems in the $\delta$ burst that after a period $t_0-t$ become observable as short period DWD in the LGWA frequency band. This shows as late periods of SFH lead to a DWD gain up to 3 times larger than early period SFH, for equal SFR. This explains why the thin disk contains the majority of the DWD systems, together with the bigger mass,  and as observed in Sect.~\ref{subsec:SFH}, the presence of a long tail in the bulge SFH results in a bigger DWD population with respect to the thick disk, while having comparable total masses.

\subsection{Extragalactic rates}
\label{app:extra_rates}
\begin{figure}
    \centering
    \includegraphics[width=1\linewidth]{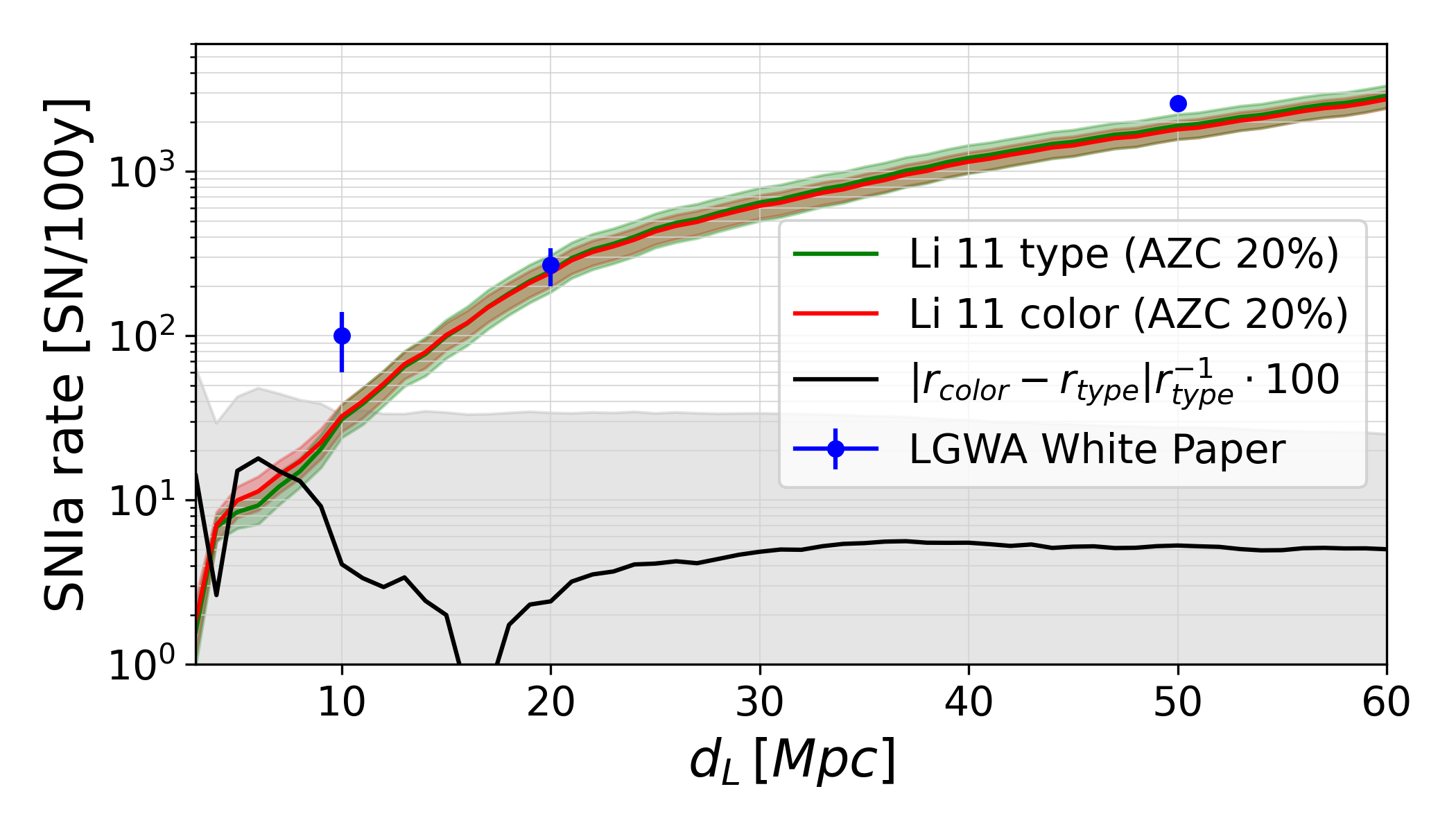}
    
    \caption{Comparison between rates obtained in this work (with 20\% avoidance zone correction factor (AZC), see Sect.~\ref{subsec:errori_popolazione}) and LGWA White Paper \citep{Ajith:2024mie}. We report the rate calculated preferably from the $B - K$ color (used in the rest of the paper, here in red) and the rate obtained using preferably the morphological type (here in green). The black line shows the discrepancy in percent between the two estimations. The error bands are obtained with Monte Carlo sampling using the statistical errors on the parameters given by \cite{SNIa_rate_Li}.}
    \label{fig:F4}
\end{figure}

As mentioned in Sect. \ref{subsec:extra_rates}, the SN Ia rate within 10\,Mpc is $31 \pm 7$ SN/100 yrs, $245 \pm 67$ SN/100 yrs for a 20\,Mpc horizon and $650 \pm 157$ SN/100 yrs for a 30\,Mpc horizon. These results, accounting for the statistical errors, are in discrete accordance with the rates reported in \cite{Ajith:2024mie}, based on an independent approach. A comparison between the rates obtained in this paper and the rates reported by \cite{Ajith:2024mie} is presented in fig.~\ref{fig:F4}. The figure presents also a comparison with the rates obtained by using preferably the morphological type instead of the color\footnote{The $B - K$ color is still used where the morphological type is not available.}. The relative discrepancy between the two measurements is shown in gray, and is compatible with 0 within the errors. We observe the rates obtained with the type to be consistently more abundant by 5\% for $d_L>20$\,Mpc.

\section{Sensitivity to initial conditions choices}
\label{appendix:initial_conditions}
In this Appendix we evaluate the sensibility of the population's characteristics with respect to the choice of the initial conditions, namely the parametrization of the distribution of $m_1$, $q=m_2/m_1$, $d$ and $\varepsilon$ for the $\delta$-burst population at zero age. 

The global abundance of super-Chandrasekhar sources is fixed by the SN Ia rate, and is thus not sensitive at all to changes in the parametrization. The abundance of sub-Chandrasekhar binaries instead varies; this population however is not detectable if extragalactic, and of secondary importance if Galactic.

The main effect of a change in the initial distributions is thus only a small rearrangement of the sources in the parameters space. In particular we consider only the change in the final mass distribution, since the frequency distribution is set by Eq. \ref{eq:densita_frequenza_espl}, and the final eccentricity is always $\varepsilon=0$.

The process leading to a formation of a close DWD is highly non-linear with respect to little variations of the parameters; however we can directly relate the final mass distribution to the initial parameter distributions in order to linearly correlate the densities in parameter space. Given an initial parameter distribution $\iota(\vartheta)$ dependent of some parameters generically called $\vartheta$ (primary mass, mass ratio, separation and eccentricity, here assumed as independent), the density in mass space is 
\begin{equation}
    \Delta_m \equiv \frac{dN(m_1, m_2)}{dm_1 dm_2}=\int_0^{t_0}dt\,\Gamma(t) \Delta_{\delta, m}(t_0-t)[\iota] 
\end{equation}
where $\Gamma(t)$ is the SFH and $\Delta_{\delta, m}(\tau)[\iota]$ is the number density in mass space of the binaries in the $\delta$-burst that after a time $\tau$ from zero age fall into the LGWA frequency band. This quantity is both a function of time for fixed initial conditions $\iota$ and a functional of the initial distributions $\iota$ once the time $\tau$ is fixed. For simplicity we used $\Delta_{\delta, m}[\iota] \equiv \int_0^{t_0}d\tau\,\Delta_{\delta, m}(\tau)[\iota]$ to estimate the errors introduced by the uncertainties in the initial parameter distributions.

\subsection{Primary IMF}

\begin{figure*}
    \centering
    \includegraphics[width=1\linewidth]{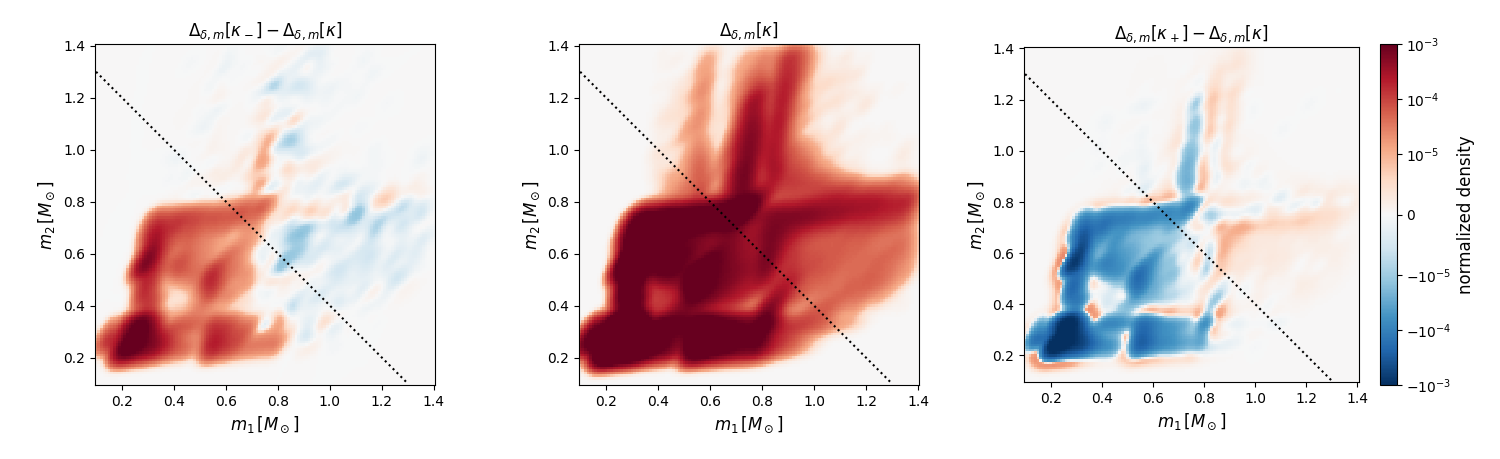}
    \caption{Comparison of the mass density distributions for extremal values of the Kroupa IMF spectral index. Center: fiducial density $\alpha=-2.7$. Left (right): difference between the density resulting from $\alpha_-=-2.76$ ($\alpha_+=-2.64$) and the fiducial density.}
    \label{fig:FC1}
\end{figure*}
The chosen primary IMF is the Kroupa IMF \citep{Kroupa:1993ga}, which in the interval $1<m_{1,i}/M_\odot<10$ is a power law with exponent $\alpha\approx-2.7$. Since the initial primary mass is directly correlated with the total mass of the resulting binary, and thus with the average S/N (see fig. \ref{fig:FA1}), among the initial conditions choices this exponent represents the most influential parameter in describing the final S/N distribution. Recent measurements found $\alpha$ in the range between 2.64 and 2.76 over a spatial scale of 800 pc \citep{KIMF_alpha}. We obtain the change in the mass distribution corresponding to these two limits by downsampling the original $\delta$-burst in order to obtain a population distributed accordingly to the new exponents. The resulting DWD mass distribution density is evaluated using a gaussian kernel density estimator and normalized to the super-Chandrasekhar fraction so that
\begin{equation*}
    1=\int_{\mathcal{D_\text{Ch}}}dm_1dm_2 \, \Delta_{\delta,m}[\kappa(\alpha)] 
\end{equation*}
where $\mathcal{D_\text{Ch}}\equiv \{(m_1, m_2)\,:\, m_1+m_2>M_\text{Ch}\}$ is the super-Chandrasekhar portion of the mass space. In fig. \ref{fig:FC1} is reported the normalized density corresponding to the fiducial Kroupa IMF distribution $\kappa(\alpha)$ (central plot), and the difference between the density distributions with the two extremal spectral indexes, $\kappa_-\equiv\kappa(\alpha_-=-2.76)$ and $\kappa_+\equiv\kappa(\alpha_+=-2.64)$, and the fiducial density distribution. The colorscale is common, and saturated for the central plot. In the super-Chandrasekhar region the difference is generally on the order of some percent of the fiducial density. In order to quantify the global discrepancy between the super-Ch densities, we introduce a mismatch measure
\[
\sigma_\pm = \sqrt{\int_{\mathcal{D_\text{Ch}}}dm_1dm_2 \Big(\Delta_{\delta,m}[\kappa_\pm]-\Delta_{\delta,m}[\kappa]\Big)^2  }
\]
Since the density is normalized, this quantity is the fractional standard deviation of the density distribution for super-Chandrasekhar DWD binaries. We obtain $\sigma_- = 5\cdot 10^{-4} $, $\sigma_+= 6\cdot 10^{-4}$, demonstrating that the error contribution due to the IMF parametrization on the entire $\mathcal{D_\text{Ch}}$ portion is negligible.

Most significantly, the variations in the marginal distribution of the chirp mass $n(\mathcal{M})\equiv\displaystyle\frac{dN}{d\mathcal{M}}$ for $m_1+m_2>M_\text{Ch}$ is limited between $\approx \pm 0.02 \, M_\odot^{-1}$ (always in normalized units, in which $N_{\text{tot}, \mathcal{D_\text{Ch}}}=1$). This corresponds to a relative error $\lesssim 2\%$ over the entire significant super-Chandrasekhar range, $1.4<M_\text{tot}/M_\odot<2.2$.

From a global point of view, we find that the fractional difference in the marginal distribution of the total mass as a function of the displacement $\delta \alpha$ from the nominal value is well represented by a linear relation
\[
\frac{\delta n_\text{tot}}{n_\text{tot}}\approx (\mathcal{M}/M_\odot-0.7) \cdot 2.1\cdot\delta \alpha
\]
We can thus see that (at first order) a displacement in the tilt of the primary IMF distribution induces a proportional tilt in the $\displaystyle\frac{\delta n(\mathcal{M})}{n(\mathcal{M})}$ marginal distribution. This discrepancy is pivoted around $\mathcal{M}_\text{piv}=0.7 M_\odot$, since this is the average value for super-Chandrasekhar systems, which are used to calibrate the abundance. A steeper IMF ($\delta\alpha<0$) will thus result in more sub-Chandrasekhar systems. The fractional difference in the marginal distribution is lower than $5\%$ over the whole observable range, and lower than $2\%$ over the super-Chandrasekhar subpopulation.

\subsection{Initial mass ratio}
We repeat the procedure described in the previous section, varying the mass ratio distribution. While we assumed a flat distribution for $q=m_2/m_1$, there is evidence that the mass ratio distribution can be expressed with a power law with exponent $\gamma$ weakly depending on the mass of the primary \citep{Duchene:2013cba}. More recently, \cite{IMF_q} found $\gamma(m_1\approx1M_\odot) = 2.12 \pm 0.19$, $\gamma(m_1\approx 1.4 M_\odot)=0.03 \pm 0.12$ and  $\gamma(m_1\approx2.4M_\odot) = -0.42 \pm 0.27$. While the decreasing trend of $\gamma$ as a function of $m_1$ is established, this measure of the spectral index for $m_1\approx 1 M_\odot$ is in tension with other estimations \citep{Duchene:2013cba}, probably due to restrictions in the considered mass ratio range; this subpopulation is however not relevant for LGWA observations since it produces the extremely light component of sub-Chandrasekhar DWD. The value for intermediate masses is compatible with a flat distribution, as assumed; finally, for higher masses $\gamma = -0.42 \pm 0.27$ is compatible with previous results.

To prove the robustness of our results, we thus evaluate the changes in the mass ratio distribution by setting $\gamma =-0.42$ on the entire mass range. We obtain a super-Chandrasekhar mismatch value (as defined in the previous section) $\sigma = 2\cdot 10^{-3}$. The marginal distribution of the chirp mass is still confined within $2\%$ of the original distribution for super-Chandrasekhar binaries, and within $5\%$ for the whole range, thus with an effect comparable with the errors relative to the uncertainties in the primary mass.

Note that the (although very limited) effect of a negative $\gamma$ is to enhance the high-mass end of the total mass spectrum, thus providing more objects with higher S/N.

\subsection{Error propagation on the S/N distribution}

Since the GW amplitude depends from $m_1$, $m_2$ through the chirp mass $\mathcal{M}$, only the deviations in the chirp mass distribution will propagate into deviations in the S/N distribution. Since the S/N depends on additional parameters ($d_L$, $\alpha$, $\delta$, $\psi$, $\theta$), local fluctuations in the $\mathcal{M}$ distribution result in "diluted" fluctuations of the S/N distribution across the entire S/N range. 

As an example, consider the effect of the luminosity distance distribution of the binaries. A (possibly big, and for example positive) deviation of the chirp mass distribution in a small interval $[\mathcal{M}^*, \mathcal{M}^*+\delta\mathcal{M}]$ would increase the number of binaries with chirp mass $\approx\mathcal{M}^*$ across the whole range of distance, both near and extremely far from the detector; thus the excess of binaries will be distributed across the entire S/N spectrum, despite being precisely localized in the chirp mass distribution. Similarly, the distributions of the other parameters contribute to smooth out local deviations of the chirp mass distribution from the fiducial one.

Under the hypothesis of a complete smoothing over the S/N range, the only effect is a global fluctuation of the number of systems; this corresponds by definition to a net zero variation for the super-Chandrasekhar population. For a more realistic partial smoothing, the S/N distribution deviations are strongly bounded by the maximum deviations in the chirp mass distribution, which are $<3\%$ for the super-Chandrasekhar population and $< 7\%$ for the sub-Chandrasekhar population, accounting for both primary IMF and mass ratio errors found before. We use these conservative upper bounds to account for the possible errors in the binary IMF of the population. A more detailed computation of the true error is not relevant, since this is not the main source of error in our population model.

\section{Fisher matrix formalism}
\label{appendix:Fisher}
In this Appendix, the Fisher matrix formalism is outlined with a practical approach in order to better understand the \textsc{GWFish} outputs.

\subsection{Fisher matrix and S/N}
Let $d_\theta(t)$ be the signal stream from the detector, that consists of a real GW signal modeled by an approximant $h_\theta(t)$ that depends on a parameter set $\{\theta_i\}$, over a stochastic Gaussian noise $n(t)$. The GW signal $h_\theta(t)$ is obtained from the tensor $h_{ij}$ by contracting it with a response tensor $\mathcal{A}^{ij}$ that represents the sensitivity of the detector. Note that the tensor $\mathcal{A}^{ij}$ is a function of time, as the detector is moving over time (with the Moon around the Earth and the Sun for LGWA). For short-time signals this variation is negligible, but since a DWD observation is obtained by integrating the signal over the entire mission lifetime this effect must be considered. The calculation of the response tensor is implemented in \textsc{GWFish}. The estimation of the errors is carried out by applying the Fisher matrix method; the likelihood function can be approximated by a multivariate Gaussian distribution:
\begin{equation}
    \mathcal{L}(d_\theta|\theta)=\mathcal{N}\exp{\Big\{-\frac{1}{2}\Delta\theta^i(\mathcal{C}^{-1})^j_i\Delta\theta_j\Big\}}
\end{equation}

where $\Delta\theta=\theta-\overline{\theta}$ and $\mathcal{C}$ is the covariance matrix; $\mathcal{C}^{-1}\equiv \mathcal{F}$ is the Fisher matrix, and can be obtained as:
\begin{equation}
     \mathcal{F}_{ij}=(\partial_ih|\partial_jh)\Big|_{\theta=\overline{\theta}}=4\Re\int_0^\infty\frac{1}{S_n(f)}\frac{\partial h}{\partial \theta_i}\frac{\partial h^*}{\partial \theta_j}\Big|_{\theta=\overline{\theta}}df
\end{equation}
where $S_n(f)$ is the noise power spectral density of the detector, assumed to be known. The Wiener dyadic product $(a|b)$ between two (complex) functions in frequency space is defined as
\begin{equation}
    (a|b)\equiv 4\Re\int_0^\infty \frac{a(f)b^*(f)}{S_n(f)}df
\end{equation}
This approach is valid only under the Gaussian approximation of the likelihood, and it could fail if the approximation is not satisfied. The validity of this method is usually compromised at very low S/N values or for deviations of the noise from Gaussianity.
The S/N in this setting is 
\begin{equation}
\label{eq:S/N_analytical}
    \frac{S}{N}\equiv\frac{(d|h)}{\sqrt{(h|h)}}\overset{\star}{=}\sqrt{4\int_0^\infty \frac{h(f)h^*(f)}{S_n(f)}df}=\sqrt{(h|h)}
\end{equation}
Where the equivalence marked with $\star$ indicates the expectancy value in the approximation of zero-mean Gaussian noise, called ``optimal S/N''.

\subsection{S/N and $\sigma_{d_l}$ relation}
\label{app:SNR_Dl_limit}
In general, the GW amplitude depends on $d_l$ as $h \propto d_l^{-1}$, thus $\partial_{d_L}h=-\frac{1}{d_L}h$. It follows that 
\begin{equation}
\label{eq:fisher_from_S/N}
\mathcal{F}_{d_Ld_L}=(\partial_{d_L}h|\partial_{d_L}h)=\frac{(h|h)}{d_L^2}=\frac{\text{S/N}^2}{d_L^2}
\end{equation}
At the same time, the error on $d_L$ is $\sigma_{d_L}=\sqrt{\mathcal{C}_{d_Ld_L}}$; the covariance matrix is obtained inverting the entire Fisher matrix $\mathcal{F}$, but for the estimation of $\sigma_{d_L}$ only the entries of $\mathcal{F}$ that in $\mathcal{C}$ show a covariance with $d_L$ are needed. In particular, the distance is strongly degenerated with the inclination angle $\theta$, so that the covariance cov$_{d_L\theta}$ can not be neglected. Inverting only the needed block results in the reduced Fisher matrix:
\begin{equation}
 \Tilde{\mathcal{F}} =  \left[ \begin{array}{cc}
 \mathcal{F}_{d_Ld_L} & \mathcal{F}_{d_L\theta} \\ \mathcal{F}_{\theta d_L} & \mathcal{F}_{\theta\theta} \end{array} \right]=\left[ \begin{array}{cc}
 \sigma_{d_L}^2 & \text{cov}(\theta, d_L) \\ \text{cov}(\theta, d_L) &  \sigma_{\theta}^2 \end{array} \right]^{-1}
\end{equation}
which leads to 
\begin{equation}
\label{eq:fisher_from_cov}
    \mathcal{F}_{d_Ld_L}=\frac{1}{\sigma_{d_L}^2}\Bigg(1-\displaystyle\frac{\text{cov}(\theta, d_L)^2}{\sigma_{d_L}^2\sigma_{\theta}^2}\Bigg)^{-1}=\frac{1}{\sigma_{d_L}^2}\cdot k
\end{equation}
where the correction to the simple relation $\mathcal{F}_{d_Ld_L}=\sigma_{d_L}^{-2}$ has been condensed in the factor $k=(1-\kappa^2)^{-1}$, where $\kappa$ is the Pearson correlation coefficient. Note that $k\geq1$ since $-1<\kappa<1$.

Matching eq. \ref{eq:fisher_from_cov} and eq. \ref{eq:fisher_from_S/N} results in
\begin{equation}
    \frac{\sigma_{d_L}}{d_L}=\frac{\sqrt{k}}{\text{S/N}}
\end{equation}

This means that with logarithmic axes (Fig. \ref{fig:F6}, \ref{fig:F7}) the population will be characterized by the strong constrain of the diagonal $\log(\text{S/N})=\log(\sigma_{d_L}/d_L)$ corresponding to null covariance, but will extend over the diagonal for higher values of covariance. The degeneracy between $d_L$ and $\theta$ can be very accentuated, so the population is expected to easily detach from the limit condition $k=0$.

\end{appendix}

\end{document}